\documentstyle[aps,epsfig,eqsecnum,twocolumn,subfigure]{revtex}
\begin{document}

\catcode`\@=11

\def\maketitle2{\par 
\begingroup
\let\cite\@bylinecite
\def\thefootnote{\fnsymbol{footnote}}%
\twocolumn[\@maketitle2\vskip2pc]%
\thispagestyle{plain}\@thanks
\endgroup
\def\thefootnote{\arabic{footnote}}%
\setcounter{footnote}{0}%
\let\maketitle2\relax \let\@maketitle2\relax
\let\@thanks\relax \let\@authoraddress\relax \let\@title\relax
\let\@date\relax \let\thanks\relax \let\@abstract\relax 
\let\@pacs\relax}

\def\abstract#1{\gdef\@abstract{{\par 
\bgroup
\ifdim\prevdepth=-1000pt \prevdepth0pt\fi
\hsize\columnwidth
\dimen0=-\prevdepth \advance\dimen0 by17.5pt \nointerlineskip
\small\vrule width 0pt height\dimen0 \relax}{~~}#1\egroup}}

\def\pacs#1{\gdef\@pacs{{\par 
\bgroup
\hsize\columnwidth \parindent0pt
\ifdim\prevdepth=-1000pt \prevdepth0pt\fi
\dimen0=-\prevdepth \advance\dimen0 by20pt\nointerlineskip
\egroup} PACS numbers:~#1}}

\def\preprint#1{\gdef\@preprint{Preprint number:~#1}}

\def\@maketitle2{
\@title
\ifdim\prevdepth=-1000pt \prevdepth0pt\fi
\@authoraddress
\@date
\begin{list}{}{\leftmargin=0.10753\textwidth \rightmargin=\leftmargin
\itemsep=1pc\partopsep=-1pc}
\item\@abstract
\item\@pacs; \@preprint

\end{list}
}

\catcode`\@=12

\newcommand{\C}{{\cal C}}
\newcommand{\ltsim}{ < \approx}

\title{Exact and Approximate Dynamics of the Quantum Mechanical $O(N)$ Model}

\author{
   Bogdan Mihaila\thanks{electronic mail:bogdan.mihaila@unh.edu}(a,b,c),
   Tara Athan\thanks{electronic mail:athan@lanl.gov}(d), 
   Fred Cooper\thanks{electronic mail:fcooper@lanl.gov}(d,e), 
   John Dawson\thanks{electronic mail:john.dawson@unh.edu}(a,f), 
   and Salman Habib\thanks{electronic mail:habib@lanl.gov}(d)
        }

\address{\ \\
   (a) Department of Physics, University of New Hampshire,
       Durham, NH 03824 \\
   (b) Theoretical Nuclear Physics Division, 
       Oak Ridge National Laboratory, Oak Ridge, TN \\
   (c) Chemistry and Physics Department,
       Coastal Carolina University, Conway, SC 29526 \\
   (d) Theoretical Division,  
       Los Alamos National Laboratory, 
       Los Alamos, NM 87545 \\
   (e) Department of Physics, Boston College, 
       Chestnut Hill, MA 02167 \\
   (f) Institute of Nuclear Theory, University of Washington,
       Box 351550, Seattle, WA 98195 
        }
\date{\today}

\abstract{We study a quantum dynamical system of $N$, $O(N)$
symmetric, nonlinear oscillators as a toy model to investigate the
systematics of a $1/N$ expansion. The closed time path (CTP) formalism
melded with an expansion in $1/N$ is used to derive time evolution
equations valid to order $1/N$ (next-to-leading order). The effective
potential is also obtained to this order and its properties are
elucidated. In order to compare theoretical predictions against
numerical solutions of the time-dependent Schr\"odinger equation, we
consider two initial conditions consistent with $O(N)$ symmetry, one
of them a quantum roll, the other a wave packet initially to one side
of the potential minimum, whose center has all coordinates equal. For
the case of the quantum roll we map out the domain of validity of the
large-$N$ expansion. We discuss unitarity violation in the $1/N$
expansion; a well-known problem faced by moment truncation
techniques. The $1/N$ results, both static and dynamic, are also
compared to those given by the Hartree variational ansatz at given
values of $N$. We conclude that late-time behavior, where nonlinear
effects are significant, is not well-described by either
approximation.}

\preprint{LAUR 00-1251}
\maketitle2

\section{Introduction}
\label{sec:Introduction}

Initial value problems in quantum field theory are of great interest
in areas such as heavy ion collisions, dynamics of phase transitions,
and early Universe physics. However, the solution of the corresponding
functional Schr\"odinger equation is essentially impossible and one is
forced to resort to approximate methods such as mean field approaches
of the Hartree type or the large $N$ expansion. The application of
variational techniques such as Hartree is limited in scope since the
errors are uncontrolled. While $1/N$ methods promise better error
control since they are based on a systematic expansion, at
next-to-leading order these methods can become extremely complicated
and expensive to implement. The motivation for our work in this paper
is to implement the $1/N$ expansion at the first nontrivial order in a
quantum mechanical example. Not only does this simplify the analysis
but it also opens the possibility of comparing the approximate results
with numerical simulations of the time dependent Schr\"odinger
equation, a luxury not available in the field theoretic case. However,
it should be kept in mind that quantum mechanics and quantum field
theory are very different. For example, in the quantum mechanics
applications discussed below, the $O(1/N)$ corrections do not
correspond to inter-particle collisions (as they do in field theory)
since we are restricting ourselves to one-particle quantum
mechanics. Nevertheless, as discussed in more detail below, quantum
mechanical examples provide excellent test-beds for key issues such as
positivity violation and late-time accuracy of the approximations.

The $O(N)$ model has been extensively employed in time independent
applications in statistical physics and quantum field theory
\cite{ref:largeN,ref:CJP} and several recent applications have studied
time-dependent phenomena. The dynamics of the chiral phase transition
following the expansion of a quark-gluon plasma produced during a
relativistic heavy ion collision has been modeled by an $O(4)$
$\sigma$-model at leading order in $1/N$\cite{ref:LDC}. The
nonequilibrium dynamics of an $O(N)$-symmetric $\lambda\phi^4$ theory,
again treated at leading order, has been investigated in detail
\cite{ref:CHKM}. Even at leading order, the $1/N$ expansion captures
the phase transition, but does not contain enough of the dynamics to
allow for rethermalization, since direct scattering first occurs at
next order. The $O(N)$ model has been used in inflationary models of
the early Universe \cite{ref:qroll} with the scalar field often
starting at the top of a hill in the potential and ``rolling'' down,
giving rise to a quantum roll problem. It has also been applied to
study primordial perturbations arising from defect models of structure
formation \cite{ntds}.

The general method for obtaining the dynamical $1/N$ approximation via
path integral techniques in quantum field theory was discussed earlier in
Ref. \cite{ref:CHKMPA} and applied later \cite{ref:PhysicaD,ref:MDC}
to a quantum mechanical system of $N+1$ coupled oscillators (a
one-dimensional version of scalar electrodynamics). Two different sets
of approximate actions were considered, which differed by terms of
order $1/N^2$, both of them being energy conserving.  The first method
in Ref.~\cite{ref:MDC} was a perturbative expansion of the generating
functional in powers of $1/N$.  The second method was to first
Legendre transform the action to order $1/N$, and then find the
equations of motion.  When these two methods diverged from each other,
they also diverged from an exact solution for the case $N=1$. However,
due to computational restrictions it was not possible to study
numerically the accuracy of the approximation as a function of
$N$. Remedying that deficiency is the main motivation of the present
study, since for the quantum roll problem numerical solutions can be
obtained for arbitrary $N$.

One of the subtle issues in expansions involving moment based
truncation schemes such as $1/N$, which is present both in quantum
field theory and in quantum mechanics, relates to the imposition of
constraints arising from the positivity of the underlying probability
density function or functional.  The importance of these constraints
is well known in areas such as turbulence and beam dynamics
\cite{real}. In this paper, we show that possible violations of these
constraints must be tamed in $1/N$ expansions if the approximation is
at all expected to succeed at moderate values of $N$. This is possible
by using certain resummation schemes which we will discuss elsewhere.

In this paper we show that the naive next-to-leading order $1/N$
expansion violates unitarity (or more generally, positivity) leading
to an instability at least for $N$ less than some value $N_T$. We have
numerical evidence for a sharp threshold at $N\sim N_T$ beyond which
we have not been able to detect an instability. This behavior appears
to be related to the nature of the effective potential at
next-to-leading order: At this order, the effective potential has the
property of not being defined everywhere for values of $N < N_{c}$,
where $N_{c}$ depends on the values of the parameters specifying the
potential, and $N_T\sim N_{c}$. However, for $N > N_{c}$, the
effective potential is defined everywhere.

A comparison of the $1/N$ expansion and Hartree is of interest since
both agree at infinite N. At finite $N$, the next-to-leading order
large N and Hartree approximations differ and provide alternative
routes to improving the leading order result which, for the quantum
roll problem, consists of harmonic oscillations in $\langle
r^2\rangle$ where $r$ is the radial degree of freedom. At finite
values of $N$, the inclusion of nonlinearities leads to amplitude
modulation effects on top of the harmonic motion. The ability to
capture this modulation is a good test for the next-to-leading order
large N and Hartree approximations. Our numerical results provide
evidence that neither of these methods are satisfactory at late times
(relative to the oscillation time), though they work reasonably well
at short to intermediate times.

Our results suggest that it is important to find ways to improve the
naive $1/N$ expansion at next-to-leading order. Work using resummation
schemes is in progress and short discussions of relevant issues are
included in this paper.

The paper is organized as follows. In Section \ref{sec:ONmodel} we
present the $O(N)$ model as it pertains to quantum mechanics and in
Section \ref{sec:LargeN} we derive equations of motion for the
large-$N$ approximation to order $1/N$. We derive the corresponding
equations for the time dependent Hartree approximation (TDHA) in
Section \ref{sec:TDHA}.  In section \ref{sec:eqtime} we show how the
same TDHA equations can be obtained from an equal-time Green's
function approach which is computationally more attractive. The
energies for the various approximations are calculated in Section
\ref{sec:energy}. Section \ref{sec:Initial} describes the two initial
conditions which preserve the $O(N)$ symmetry, namely a quantum roll,
and the time evolution of an offset Gaussian centered at an $O(N)$
symmetric point. In Section \ref{sec:effpot} we determine the
effective potential to both order $1/N$ and for the Hartree
approximation. Numerical results and comparisons with the
approximations are discussed in Section \ref{sec:Res} and our
conclusions are discussed in Section \ref{sec:Cons}.

\section{The $O(N)$ model}
\label{sec:ONmodel}

The Lagrangian for the $O(N)$ model in quantum mechanics is given by:
\begin{equation}
   L(x,\dot{x}) 
   = 
   \frac{1}{2} \sum_{i=1}^N \dot{x}_i^2 - V(r)  \>,
   \label{eq:Lagmod}
\end{equation}
where $V(x)$ is a potential of the form
\begin{equation}
   V(r)  =  \frac{g}{8 N} \, 
      \left  ( 
         r^2  -  r_0^2 
      \right )^2
   \>,  \qquad
   r^2 = \sum_{i=1}^N x_i^2 \>.
   \label{eq:classi}
\end{equation}
The time-dependent Schr\"odinger equation for this problem is given
by: 
\begin{equation}
i \, \frac{\partial \psi(x,t)}{\partial t} \>=
\left  \{ - \frac{1}{2} \sum_{i=1}^N \frac{\partial^2}{\partial x_i^2}
+ V (r) \right \} \, \psi(x,t).
\end{equation}
For arbitrary initial conditions, given present computational
constraints, these equations can be numerically integrated only for
small $N \le 4$.  The initial conditions for the quantum roll problem
allow a numerical solution for all $N$, and in this case we can
attempt to study fully the behavior of the large-$N$ expansion. (For
the shifted Gaussian initial conditions, however, this is not
possible, and we used numerical solutions obtained for $N=1$ and $2$
to benchmark the large-$N$ approximations and the TDHA solutions at
short times.)

The symmetry of the quantum roll problem is such that only the radial
part of the wave function is of interest. Assuming a solution of the
form
\begin{equation}
   \psi(r,t) = r^{(1-N)/2} \phi(r,t),
\end{equation} 
the time dependent Schr\"odinger equation for $\phi(r,t)$ reduces
to\cite{ref:BlazotRipka}: 
\begin{equation}
   i \, {\partial \phi(r,t) \over \partial t} \>=\left \{ -
      \frac{1}{2}\frac{\partial^2}{\partial r^2}  
      + U(r) \right \} \, \phi(r,t)
   \label{eq:redham}
\end{equation}
with an effective one dimensional potential $U(r)$ given by
\begin{equation}
   U(r) 
   = 
   \frac{(N-1)(N-3)}{8 \, r^2} 
   +
   \frac{g}{8N} \, \left ( r^2 - r_0^2 \right )^2
   \>. 
\label{eq:Uofr}
\end{equation}
It is further useful to make the rescaling:
\begin{equation} 
   r^2 = N y^2 \>, \qquad r_0^2 = N y_0^2 \>.
   \label{eq:yscaling}
\end{equation}
The potential (\ref{eq:Uofr}) then becomes:
\begin{equation}
   u(y,N) 
   = 
   \frac{U(y)}{N}
   =
   \frac{(N-1)(N-3)}{8 N^2 \, y^2} + 
   \frac{g}{8} \, ( y^2 - y_0^2 )^2  \>,
   \label{eq:Uscaled}
\end{equation}
corresponding to the new Schr\"odinger equation,
\begin{equation}
   i \, {\partial \phi(y,\tilde{t}) \over \partial \tilde{t}} \>=
\left \{   
      - \frac{1}{2N^2} 
      \frac{\partial^2}{\partial y^2} 
      + u(y,N) 
   \right \} \, \phi(y,\tilde{t})
   \label{eq:schroii}
\end{equation}
where $\tilde{t} = N \, t$.
  
The method of choice to investigate the long-time behavior of the
exact solution is the split-operator method, which has been presented
in detail in Ref. \cite{ref:sp_op}. The wave function is expanded as a
Fourier series in the radial component, and the solution is obtained
as the repeated application of a time-evolution operator in
symmetrically split form. As a result, the use of a Fast-Fourier
Transform algorithm is required. For the purpose of the present
implementation, 256 radial grid points, a value of 20 for the radial
grid boundary, and a time step size of 0.01, provides a conservation
of the wave function unitarity to better than 9 significant
figures. The accuracy of the method has been established by comparing
results with a second method, where we first solve for the eigenvalues
and eigenfunctions, and then use the expansion
\begin{equation}
   \phi(r,t) 
   = 
   \ \sum_n  \, C_n \, e^{-iE_n t} \, \phi_n(r) \>,
\end{equation}
where $C_n$ was determined from the initial conditions.  This method
was restricted to moderate values of $N$. Results from the two methods
agreed in the cases where they were used together.

\section{The Large-$N$ approximation}
\label{sec:LargeN}

The large $N$ approximation has been worked out for the $O(N)$ model
in $1+3$ dimensions in Ref.~\cite{ref:CHKMPA}.  The Lagrangian
(\ref{eq:Lagmod}) with the potential function (\ref{eq:classi}) is
obtained from that paper by specializing to $0+1$ dimensions, and
replacing $\phi_a(t) \rightarrow x_i(t)$, $v \rightarrow r_0$, and
$\lambda \rightarrow g$.

To implement the large $N$ expansion, it is useful\cite{ref:CJP} to
rewrite the Lagrangian in terms of the composite field $\chi$ by
adding a constraint term to (\ref{eq:Lagmod}), given by:
\begin{equation}
   \frac{N}{2 g} \, 
      \left  [ 
         \chi - 
         \frac{g}{2 N} ( r^2 - r_0^2 ) 
      \right ]^2  \>,
\end{equation}
which yields an equivalent Lagrangian,
\begin{equation}
   L'(x,\dot{x},\chi) =
      \sum_i  
         \frac{1}{2} \left  ( 
                        \dot{x}_i^2 - \chi x_i^2 
                     \right )
       + \frac{r_0^2 }{2} \chi
       + \frac{N}{2 g} \chi^2 \>.
\label{eq:LLN}
\end{equation}
The generating function $Z[j,J]$ is given by the path integral over
the classical fields $x_i(t)$:
\begin{eqnarray*}
   Z[j,J] 
   = e^{iW[j,J]} 
   &=&  
   \int {\rm d}\chi \ \prod_i {\rm d} x_i \, 
        \exp
        \Bigl \{ 
           i \, S[x,\chi;j,J] 
        \Bigr \}  \>,
   \\
   S[x,\chi;j,J]
   &=&  
    \int_{\C} {\rm d}t \, 
    \Bigl \{  
       L' + \sum_i j_i x_i + J \chi  
    \Bigr \}  \>.
\end{eqnarray*}
The effective action, to order $1/N$, is obtained by integrating the
path integral for the generating functional for the Lagrangian
(\ref{eq:LLN}), over the $x_i$ variables, and approximating the
integral over $\chi$ by the method of steepest descent (keeping terms
up to order $1/N$).  A Legendre transform of the resulting generating
functional then yields the effective action, which we find to be:
\begin{eqnarray}
   \lefteqn{\Gamma[q,\chi] = }
   \nonumber \\
   &&
   {\displaystyle \int_{\C} {\rm d}t} \, 
       \biggl \{ 
          \frac{1}{2} \sum_i \,
          \Bigl [ 
             \dot{q}_i^2(t) - \chi(t) \,  q_i^2(t) 
          \Bigr ] 
          +
          \frac{i}{2} \sum_i \,
          \ln \, [ G_{ii}^{-1}(t,t) ]
   \nonumber \\
   && \qquad
          + \frac{r_0^2}{2} \, \chi(t)
          + \frac{N}{2 g} \, \chi^2(t)
          + \frac{i}{2} \ln [ D^{-1}(t,t) ]
       \biggr\}
   \>, 
\label{eq:effaction}
\end{eqnarray}
where the integral is over the close time path $\C$, discussed in
Ref.~\cite{ref:CHKMPA} and $q(t) = \langle x_i(t) \rangle$.  Here
$G^{-1}_{ij}(t,t')$ and $D^{-1}(t,t')$ are the lowest order in $1/N$
inverse propagators for $x_i$ and $\chi$, given by
\begin{eqnarray*}
   G^{-1}_{ij}(t,t') 
   & = &
      \left  \{
         \frac{ {\rm d}^2 }{ {\rm d} t^2 } + \chi(t)
      \right \} \, \delta_{\C} (t,t') \, \delta_{ij} 
   \equiv  
   G^{-1}(t,t') \, \delta_{ij} 
   \>, \\
   D^{-1}(t,t') 
   & = &
      - \frac{N}{g} \delta_{\C}(t,t')
      -  \Pi(t,t')
   \>,
\end{eqnarray*}
where
\begin{eqnarray}
   \Pi(t,t') 
   & = &
      - \frac{i}{2} \, \sum_{i,j} G_{ij}(t,t') \, G_{ji}(t',t)
   \nonumber \\ && 
      + \sum_{i,j} \, q_i(t) \, G_{ij}(t,t') \, q_j(t') \>.
   \label{eq:PilrgNdef}
\end{eqnarray}
Here $\delta_{\C} (t,t')$ is the closed time path delta function.  

The equations of motion for the classical fields $q_i(t)$, to order
$1/N$, are 
\begin{eqnarray}
   &&
   \left  \{
      \frac{{\rm d}^2 }
           {{\rm d} t^2 }
      + \chi(t)
   \right\} q_i(t) 
   \nonumber \\ 
   &&
   + i \, \sum_{j} \int_{\C} {\rm d}t' \,
         G_{ij}(t,t') \, D(t,t') \, q_j(t')
   = 0 
   \>,
\label{eq:phieom}
\end{eqnarray}
with the gap equation for $\chi(t)$ given by
\begin{equation}
   \chi(t)
   =
   - \frac{g}{2N} \, r_0^2 + 
     \frac{g}{2 N} \sum_i
      \left [
         q_i^2(t) \, + \, \frac{1}{i} \, {\cal G}_{ii}^{(2)}(t,t)
      \right ]
   \>.
\label{eq:Chieqn}
\end{equation}
The next-to-leading order $x_i$ propagator ${\cal G}_{ij}^{(2)}(t,t')$
and self energy $\Sigma_{ij}(t,t')$ to order $1/N$ turn out to be
\begin{eqnarray}
   &&
   {\cal G}_{ij}^{(2)}(t,t')
   \ = \
   G_{ij}(t,t') \label{eq:Gfull}  \\
   &&
   - \, \sum_{k,l}
      \int_{\C} {\rm d}t_1 \, \int_{\C} {\rm d}t_2 \,
      G_{ik}(t,t_1) \, \Sigma_{kl}(t_1,t_2) \, G_{lj}(t_2,t') \>,
   \nonumber \\
   &&
   \Sigma_{kl}(t,t')
   \ = \
   i \, G_{kl}(t,t') \, D(t,t')
     - q_k(t) \, D(t,t') \, q_l(t')  
   \>.
   \nonumber
\end{eqnarray}
These equations agree with (2.18--2.22) of Ref.~\cite{ref:CHKMPA}. We
mention here that the actual equation for $\cal G$ which follows from
the effective action differs from Eq.~(\ref{eq:Gfull}) in that the
final $G$ in the integral equation is replaced by the full $\cal G$.
This leads to a partial resummation of the $1/N$ corrections which
which guarantees positivity of $\langle x^2(t)\rangle$ (this
restricted result does not imply that the full positivity problem for
the density matrix has been solved). However, it does not improve the
long-time accuracy of the results \cite{ref:MDCwork}.

In order to solve for $D(t,t')$, we first write
\begin{equation}
   \frac{N}{g} \, D(t,t') \ = \
      - \, \delta_{\C}(t,t')
      \, + \, 
      \frac{N}{g} \, \Delta D(t,t')
   \>.
\label{eq:Dsubst}
\end{equation}
Then $\Delta D(t,t')$ satisfies the integral equation,
\begin{equation}
   \frac{N}{g} \, \Delta D(t,t')
   \ = \
   \frac{g}{N} \, \Pi(t,t')
   - \int_{\C} {\rm d}t'' \,
            \Pi(t,t'') \, \Delta D(t'',t')
   \>,
\label{eq:Dtileqn}
\end{equation}
in agreement with (2.13--2.16) of Ref.~\cite{ref:CHKMPA}.  

We are now in a position to solve these coupled equations for the
motion of $q_i(t)$ and $\chi(t)$ for given initial conditions.  For
the initial conditions discussed in Sec.~\ref{sec:Initial}, we find:
\begin{equation}
   G_{ij}(t,t')/i
   =
   \delta_{ij} \, f(t) f^{\ast}(t') \>,
   \label{eq:Gintoff}
\end{equation}
where $f(t)$ and $f^{\ast}(t)$ satisfy the homogeneous equation,
\begin{equation}
   \left  \{
      \frac{{\rm d}^2 }
           {{\rm d} t^2 }
      + \chi(t)
   \right\}
   \left  (
      \begin{array}{c}
         f(t) \\
         f^{\ast}(t)
      \end{array}
   \right )
   = 0  \>,
   \label{eq:ffstareq}
\end{equation}
with initial conditions:
\begin{equation}
   f(0) =\sqrt{G}  \>, 
   \qquad\qquad 
   \dot{f}(0) = 1/( 2 \sqrt{G} ) \>.
   \label{eq:f0dotf0}
\end{equation}
However, $\Delta D(t,t')$ cannot be factored into products of
functions like $G_{ij}(t,t')$.

We solve (\ref{eq:phieom}) and (\ref{eq:Chieqn}) simultaneously with
(\ref{eq:Gfull}) and (\ref{eq:Dtileqn}), using the Chebyshev expansion
technique\cite{ref:BMIM} of Appendices A and B of Ref.~\cite{ref:MDC}.

\section{The time dependent Hartree approximation}
\label{sec:TDHA}

It is useful to compare our results for the large-$N$ approximation to
the time dependent Hartree approximation (TDHA) suitably formulated
for the $O(N)$ problem. The static Hartree approximation is based on
the idea of varying the parameters of a Gaussian wave function so as
to minimize the energy (the generalization to the time-dependent case
is given below). For the $O(N)$ problem this amounts to placing an
$N$-dimensional Gaussian some (radial) distance away from the origin
and then carrying out the minimization procedure. In contrast, the
leading-order large $N$ wave function is a Gaussian which is locked at
the origin. At infinite $N$ the TDHA becomes exact and equivalent to
the leading order large-$N$ approximation, a well-known result. At
finite $N$, the TDHA and the next-to-leading order large $N$
approximation can be thought of as two competing schemes to improve on
the leading-order result.

There are several ways of implementing the Hartree approximation: The
most common is by using the time-dependent variational principle of
Dirac \cite{ref:Dirac,ref:notDirac,ref:CDHR}.  This has the advantage
of giving a classical Hamiltonian description for the dynamics of the
variational parameters, which can be hidden in other formulations.

The idea behind this approach is that the variation of
\begin{eqnarray}
   \Gamma[\psi,\psi^{\ast}] & = & 
   \int {\rm d}t \,  
   \langle \psi(t) \, | \, 
      i \frac{\partial}{\partial t} - H  \, 
   | \, \psi(t) \rangle
\end{eqnarray}
is stationary for the exact solution of the Schr\"odinger equation.
We consider Gaussian trial wave functions of the form:
\begin{eqnarray}
   && \psi(x, t)  = \
   {\cal N} \
   \exp \,
   \biggl  [
      i p_i(t) \, z_i(t)
   \label{eq:trial} \\ 
   &&
   - z_i(t)
        \left ( \frac{G^{-1}_{ij}(t)}{4} - \Pi_{ij}(t) \right )
     z_j(t)
    \biggr ]
   \>,
   \nonumber 
\end{eqnarray}
where ${\cal N}$ is the normalization constant, and we have set
$z_i(t) = x_i - q_i(t)$.  Here $q_i(t)$, $p_i(t)$, $G_{ij}(t)$ and
$\Pi_{ij}(t)$ are time-dependent variational parameters, to be
determined by minimizing the Dirac action. We note that $\Pi_{ij}(t)$,
which is used only in this section, is conjugate to $G_{ij}(t)$ and is
not to be confused with the self energy $\Pi(t,t')$ defined in
Eq.~(\ref{eq:PilrgNdef}).

The $n$-point functions can be calculated from the generating
functional using the formula,
\begin{equation}
   \langle 
      z_i z_j \cdots z_n
   \rangle
   =
   \frac{\partial^n Z[j] }
        {\partial j_i \partial j_j \cdots \partial j_n}
   \biggr |_{j = 0}
   \>,
   \label{eq:NptZ}
\end{equation}
where
\begin{eqnarray}
   Z[j]
   & = &
   {\cal N}^2 \
   \int \
      \prod_s {\rm d} x_s 
   \nonumber \\ 
   && \times 
      \exp \biggl [ - \, \frac{1}{2} \,
                     z_i(t) G^{-1}_{ij}
                     z_j(t)
                   \, + \,
                   j_i \, z_i(t)
           \biggr ]
   \nonumber \\
   & = &
   \exp \left [{ j_i G_{ij} j_j \over 2} \right ]
   \>.
\end{eqnarray}
The expectation value of the time derivative is given by
\begin{eqnarray}
   \left  \langle 
      i \frac{\partial}{\partial t}
   \right \rangle
   \ = \
   p_i \dot{q}_i
   \, - \,
   G_{ij} \dot{\Pi}_{ij}
\end{eqnarray}
and the expectation value of the kinetic energy is
\begin{equation}
   \left  \langle 
      - {1 \over 2} \,
        \frac{\partial^2}{\partial x_i^2}
   \right \rangle
   \ = \ {p_i p_i  \over 2}
   \, + \,
   \frac{1}{8} \,  G^{-1}_{ii}
   \, + \,
   2 \, \Pi_{ij} G_{jk} \Pi_{ki}
   \>.
\label{eq:kin_erg}   
\end{equation}
For the expectation value of $V$ we first expand the potential
in a Taylor series about $z_i=0$,
\[
   V(q,z) 
   =  
   V(q) + V_i(q) \, z_i + 
   \frac{1}{2} V_{ij}(q) \, z_i z_j + \ldots
\]
where
\begin{eqnarray}
   V_{i}(q)
   & = &
   \frac{g}{2N} \
   q_i \, \left ( q_s q_s - r_0^2 \right )
   \>,
   \label{eq:Vee} \\
   V_{ij}(q)
   & = &
   \frac{g}{2N} \,
   \left [
      \delta_{ij} \left ( q_s q_s - r_0^2 \right )
      \, + 2 \, q_i q_j
   \right ]
   \>,
   \nonumber \\
   V_{ijk}(q)
   & = &
   \frac{g}{N} \,
   \left (
      \delta_{ij} q_k + \delta_{ik} q_j + \delta_{jk} q_i
   \right )
   \>,
   \nonumber \\
   V_{ijkl}(q)
   & = &
   \frac{g}{N} \,
   \left (
      \delta_{ij} \delta_{kl}
      + \delta_{il} \delta_{jk}
      + \delta_{ik} \delta_{jl}
   \right )
   \>.
   \nonumber
\end{eqnarray}
Thus 
\begin{eqnarray*}
   &&
   V(q,z) 
   =  
   \frac{g}{8N} 
   [ (q_j q_j - r_0^2)^2 + (z_i z_i)^2 + 4 (z_i z_i)
       (z_j q_j) 
   \\
   &&
   {}+ 4 (z_i q_i)^2 + 2 (z_i z_i) (q_j q_j - r_0^2) + 
       4 (z_i q_i) (q_j q_j - r_0^2)] 
\end{eqnarray*}
Taking the expectation value, we obtain:
\begin{eqnarray}
   \langle V \rangle 
   & = &
   \frac{g}{8 N} 
   [ (q_j q_j - r_0^2)^2 + 2 G_{ii}(q_j q_j - r_0^2 ) 
   \nonumber \\
   &&
   {}+ 4 G_{ij}q_i q_j + G_{ii} G_{jj} + 2 G_{ij}G_{ji} ]
   \label{eq:expcVHartree}
\end{eqnarray}
The Hartree equations of motion are Hamilton's equations for the
variational parameters:
\begin{eqnarray}
   \dot{q}_i & = & p_i
   \>,
   \label{eq:Hartreeq} \\
   \dot{p}_i
   & = &
   - V_i - \frac{1}{2} V_{ijk} G_{jk}
   \>,
   \nonumber \\
   \dot{G}_{ij}
   & = &
   2 \,
   \left ( G_{ik} \Pi_{kj}
   \, + \, G_{jk} \Pi_{ki} \right ),
   \nonumber\\
   \dot{\Pi}_{ij}
   & = &
      \frac{1}{8} \, G^{-1}_{ik} G^{-1}_{kj}
      - 2 \, \Pi_{ik} \Pi_{kj} 
      - \frac{1}{2} V_{ij}
      - \frac{1}{4} \, V_{ijkl} G_{kl}  \>.  
   \nonumber 
\end{eqnarray}
Solutions of this set of equations determine the time-dependent
Hartree approximation to the true solution of the Schr\"odinger
equation.

\section{Method of equal-time Green's functions}
\label{sec:eqtime}

Solutions of the TDHA equation (\ref{eq:Hartreeq}) require computing
the matrix inverse of $G_{ij}$. This can be difficult to carry out in
practice for large $N$. Fortunately, the method of equal-time Green's
functions provides a way to avoid this technical difficulty
\cite{ref:equalt,ref:Wetterich}. We begin by considering the
time-evolution of the one point functions:
\begin{eqnarray}
   \dot{q}_i & = & p_i
   \>,
   \label{eq:ETGFqp} \\
   \dot{p}_i
   & = &
   - V_i - \frac{1}{2} \, V_{ijk} \, G_{jk} \>,
   \nonumber \\
   & = & 
   - \frac{g}{2N} \, 
   \left \{
      q_i ( q_k q_k + G_{kk} - r_0^2 ) + 
      q_k ( G_{ik} + G_{ki} ) 
   \right \}
   \nonumber 
\end{eqnarray}
where $V_i$ and $V_{ijk}$ are given by (\ref{eq:Vee}), as well as the
evolution of the two-point functions: 
\begin{eqnarray}
   G_{ij}(t)
   & = &
   \langle z_i \, z_j \rangle
   \>, \qquad
   K_{ij}
   = 
   \langle \dot{z}_i \, \dot{z}_j \rangle \>, 
   \nonumber \\
   F_{ij}
   & = &
   \frac{1}{2} 
   \langle 
      \, [ \, z_i \dot{z}_j + \dot{z}_j z_i \, ] \, 
   \rangle
   \>.
   \label{eq:ETGFdefGFK}
\end{eqnarray}
Here we have again set $z_i(t) = x_i - q_i(t)$.  All of the
expectation values are taken with respect to the Gaussian trial wave
function, Eq.~(\ref{eq:trial}). To obtain the equations of motion for
the two-point functions, we use the exact equation of motion and the
factorization resulting from the Gaussian approximation. This yields
\begin{eqnarray}
   \dot{G}_{ij}
   & = &
   F_{ij} + F_{ji}
   \>,
   \label{eq:ETGFdotG} \\
   \dot{F}_{ij}
   & = &
   K_{ij} -
   \Bigl\langle 
      z_i \, \frac{\partial V}{\partial x_j} 
   \Bigr\rangle 
   \>,
   \label{eq:ETGFdotF} \\
   \dot K_{ij}
   & = &
   - 
   \Bigl\langle \, 
      \Bigl [
      \frac{\partial V}{\partial x_i} \, \dot{z}_j \,  
      - 
      \dot{z}_i \, \frac{\partial V}{\partial x_j} 
      \Bigr ] 
   \Bigr\rangle
   \>.
   \label{eq:ETGFdotK}
\end{eqnarray}
Here we have used the Lagrange equations of motion
\begin{equation}
   \ddot{x}_i \ + \ \frac{\partial V}{\partial x_i}
   = 0  \>,
\end{equation}
where
\begin{eqnarray}
   \frac{\partial V}{\partial x_i}
   & = &
   V_{ij} \, z_j + 
   \frac{1}{6} \, V_{ijkl} \, z_j \, z_k \, z_l 
   \nonumber \\
   && {}+ \, \text{terms with even powers of $z_i$}
   \>,
\end{eqnarray}
and the fact that for our Gaussian wave packet, $\langle z_i \rangle =
0  \>.$ The canonical commutation relations give:
\begin{equation} 
   \langle z_i \dot z_j - \dot z_j z_i \rangle 
   =
   \langle x_i \dot x_j - \dot x_j x_i \rangle 
   = i \delta_{ij}  \>, 
\end{equation}
and we have
\begin{eqnarray}
   &&
   \langle
       z_i \, z_j \, z_k \, z_l
   \rangle
   = 
        G_{ij} G_{kl}
      + G_{il} G_{jk}
      + G_{ik} G_{jl}
   \>,
   \\
   &&
   \langle
       z_i \, z_j \, z_k \, \dot{z}_l
   \rangle
   = 
        G_{ij} F_{kl}
      + F_{il} G_{jk}
      + G_{ik} F_{jl}
   \nonumber \\ && \qquad
   {}+ i
   \left (
        G_{ij}  \delta_{kl}
      + G_{jk}  \delta_{il} 
      + G_{ik}  \delta_{jl}
   \right ) / 2           
   \>,
   \\
   &&
   \langle
       \dot{z}_i \, z_j \, z_k \, z_l 
   \rangle
   = 
        F_{ji} G_{kl}
      + F_{li} G_{jk}
      + F_{ki} G_{jl}
   \nonumber \\ && \qquad
   {}- i
   \left (
        G_{kl}  \delta_{ji}
      + G_{jk}  \delta_{li}
      + G_{jl}  \delta_{ki} 
   \right )  / 2              
   \>.
\end{eqnarray}
Finally, from Eqs.~(\ref{eq:ETGFdotF}) and (\ref{eq:ETGFdotK}) we get:
\begin{eqnarray}
   \dot F_{ij}
   & = &
   K_{ij} - V_{jk} G_{ik}
   \label{eq:ETGFdotFi} \\ &&
   {}- V_{j k l m}
   \left (                
        G_{i k} G_{l m}
      + G_{i m} G_{k l}
      + G_{i l} G_{k m}
   \right ) / 6 
   \>,
   \nonumber \\
   \dot K_{ij}
   & = &
      - V_{i k} F_{k j}
      - V_{j k} F_{k i}
   \label{eq:ETGFdotKi} \\ &&
   {}- V_{i k l m}
   \left (
        G_{k l} F_{m j}
      + G_{l m} F_{k j} 
      + G_{k m} F_{l j}
   \right ) / 6
   \nonumber \\ 
   &&
   {}- V_{j k l m}
   \left (
        F_{k i} G_{l m}
      + F_{m i} G_{k l}
      + F_{l i} G_{k m}
   \right ) / 6 
   \>.
   \nonumber
\end{eqnarray}

For Gaussian initial conditions, the equal-time Green's function
method is assured to give the same result as the Hartree method, if
$F_{ij}(0)$ and $K_{ij}(0)$ satisfy the requirements:
\[
   F_{ij}(0) = 0 \>,
   \qquad \qquad
   K_{ij}(0) = G_{ij}^{-1}(0) / 4  \>.
\]
Choosing $K_{ij}$ independently of $G_{ij}$, corresponds to a mixed
initial density matrix, rather than a pure state.  If we further
choose $G_{ij}(0)$ to be diagonal and equal to the same number $G_0$,
\[ 
   G_{ij}(0) = \delta_{ij} \, G_0 \>,
\]
then $K_{ij}(0)$ is given by:
\[  
   K_{ij}(0) = \delta_{ij} / ( 4 \, G_0 ) \>.
\]
For the initial conditions pertinent to the quantum roll, $q_i(t) = 0$
for all $t$.  Then $G_{ij}(t)$, $F_{ij}(t)$, and $K_{ij}(t)$ are all
proportional to the unit matrix, and have no off-diagonal terms.  For
the offset initial condition, we choose large-$N$ symmetric initial
conditions so that $q_i(0) = q_0$, and $p_i(0) = 0$, and $G_{ij}(0) =
G_0 \delta_{ij}$.  In that case, all the $q$'s and $p$'s are
identical,
\[  
   q_i(t) = q(t) \>, \qquad p_i(t) = p(t)  \>,
\]
and the matrices $G_{ij}(t)$, $F_{ij}(t)$ and $K_{ij}(t)$ become
off-diagonal in a simple way so that all the diagonal elements are
equal and all the off-diagonal elements are equal.  That is, we can
write
\begin{eqnarray*}
   G_{ij}(t) 
   & = &
   G(t) \delta_{ij} + \bar{G}(t) ( 1 - \delta_{ij} ),
   \\
   F_{ij}(t) 
   & = &
   F(t) \delta_{ij} + \bar{F}(t) ( 1 - \delta_{ij} ),
   \\
   K_{ij}(t) 
   & = &
   K(t) \delta_{ij} + \bar{K}(t) ( 1 - \delta_{ij} ).
\end{eqnarray*}
For this case, Eqs.~(\ref{eq:ETGFqp}), (\ref{eq:ETGFdotG}),
(\ref{eq:ETGFdotFi}), and (\ref{eq:ETGFdotKi}) simplify to the
following set of coupled equations:
\begin{eqnarray}
   \dot{q} 
   & = & 
   p  \>, 
   \label{eq:etgfii} \\
   \dot{p}
   & = & 
   - \frac{g}{2N} \, q \, 
      \biggl \{ 
          N q^2 - r_0^2 + 
         (N + 2) G + 
         2 (N - 1) \bar{G}
      \biggr \} \>, 
    \nonumber \\
   \dot{G} 
   & = & 
   2 F  \>, 
   \qquad 
   \dot{\bar{G}}
   = 
   2 \dot{\bar{F}}  \>, 
   \nonumber \\
   \dot{F}
   & = & 
   K - \frac{g}{2N} 
      \biggl \{
         G \, [ \, (N +2) (q^2 + G) - r_0^2 \, ] 
   \nonumber \\ 
   && \qquad 
         {}+ 2 (N - 1) \, \bar{G} \, (q^2 + \bar{G})
      \biggr \}  \>, 
   \nonumber \\
   \dot{\bar{F}}
   & = & 
   \bar{K} - \frac{g}{2N} 
      \biggl \{
         \bar{G} \, 
            [ \, (3N - 2) \, q^2 + (N + 4) \, G 
   \nonumber \\ 
   && \qquad
              {}+ 2(N - 2) \, \bar{G} - r_0^2 \, 
            ] + 
         2 G \, q^2  
      \biggr \}  \>, 
   \nonumber \\
   \dot{K}
   & = & 
   - \frac{g}{N} 
      \biggl \{
         F \, [ \, 
            (N + 2) (q^2 + G) - r_0^2 \, 
              ] +
   \nonumber \\ 
   && \qquad
         {}+ \bar{F} \, 2 (N -1) \, (q^2 + \bar{G})
      \biggr \}  \>, 
   \nonumber \\      
   \dot{\bar{K}}
   & = & 
   - \frac{g}{N} 
      \biggl \{
         \bar{F} \, [ \, 
            (3N - 2) q^2 + (N + 2) \, G 
   \nonumber \\ 
   && \qquad
            {}+ 2(N - 2) \, \bar{G}
         - r_0^2 \, ]
         + F \, 2 (q^2 + \bar{G} ) 
      \biggr \}  \>.
   \nonumber    
\end{eqnarray}
If we let $r_0^2 = N y_0^2$, and then take the limit $N \rightarrow
\infty$, we recover the leading order in the large-$N$ result, as
discussed in Ref.~\cite{ref:equalt}.

The equal time Green's function method is easier to implement
numerically than the Hamiltonian system described by
Eq. (\ref{eq:Hartreeq}), since no matrix inversion is involved.
However, if one wants to find the wave function or the energy, instead
of just obtaining the Green's functions, matrix inversion is once
again required.

\section{Energy}
\label{sec:energy}

It is important to note that even though the Hartree and large $N$
approximations are truncations of the true dynamics, they are
nevertheless energy conserving.  In the large-$N$ approximation, to
order $1/N$, the expectation value of the Hamiltonian is given by
\begin{eqnarray}
   E
   & = &
   \frac{1}{2} \sum_i
       \left [
          \langle \dot{x_i}^2(t) \rangle + 
          \langle \hat\chi(t) \, x_i^2(t) \rangle
       \right ]  
   \label{eq:energy} \\
  && \qquad
   - \frac{r_0^2}{2} \, \langle \hat\chi(t) \rangle
   - \frac{N}{2 g} \, \langle \hat\chi^2(t) \rangle
   \>. \nonumber
\end{eqnarray}
We write these expectation values in terms of the CTP Green's
functions.  By definition, the disconnected two-point Green's
functions are introduced as
\begin{eqnarray}
   D_{\rm{dis}}(t,t')
   & = & i \langle \,
         {\rm T}_{\C}[ \hat\chi(t) \hat\chi(t')] \, \rangle
   \nonumber \\
   & = &
   i \, \chi(t) \chi(t') + {\cal D}(t,t')
   \>,
   \\
   G_{ij, \, \rm{dis}}(t,t')
   & = & i \langle \,
         {\rm T}_{\C}[ x_i(t) x_j(t')] \, \rangle
   \nonumber \\
   & = &
   i \, q_i(t) q_j(t') + {\cal G}_{ij}(t,t')
   \>,
\end{eqnarray}
where ${\cal D}$ and ${\cal G}_{a b}$ denote the {\em connected}
two-point Green's functions
\begin{eqnarray}
   {\cal D}(t,t')
   & = & \left [ 
           \frac{ \delta^2 W[J,j] }
                { \delta J(t) \, \delta J(t') }
         \right ]_{J, \, j = 0}
   \>,
   \nonumber \\
   {\cal G}_{ij}(t,t')
   & = & \left [ 
           \frac{ \delta^2 W[J,j] }
                { \delta j_i(t) \, \delta j_j(t') } 
         \right ]_{J, \, j = 0}
   \>.
\end{eqnarray}
To obtain the energy from (\ref{eq:energy}), we require the expectation
values
\begin{eqnarray*}
   \langle \hat{\chi}(t) \rangle
   & = &
   \chi(t)
   \>,
   \\
   \langle \hat{\chi}^2(t) \rangle
   & = &
   \chi^2(t) + D(t,t)/i
   \>,
   \\
   \langle x_i^2(t) \rangle
   & = &
   q_i^2(t) + {\cal G}_{ii}(t,t) / i
   \>,
   \\
   \langle \dot{x}_i^2(t) \rangle
   & = &
   \dot{q}_i^2(t) +
   \left . 
      \frac{\partial^2 \, {\cal G}_{ii}(t,t')/i}
           {\partial t \, \partial t'}
   \right |_{t=t'}
   \>,
   \\
   \langle \hat\chi(t) \, x_i^2(t) \rangle
   & = &
   \chi(t) \, \left [
                  q_i^2(t) + {\cal G}_{ii}(t,t) / i
              \right ]
   - K_{ii}(t,t,t)
   \>,
\end{eqnarray*}
where $ K_{ij}(t_1,t_2,t_3)$ is the 3-point Green's function defined
as
\begin{eqnarray*}
   K_{ij}(t_1,t_2,t_3)
   = 
   - \int_{\C} {\rm d}t \,
           G_{ik}(t_1,t) \, G_{kj}(t_2,t) \, D(t,t_3)
   \>.
\end{eqnarray*}
The energy for the next-to-leading order large-N approximation is then given
by:
\begin{eqnarray}
   E
   & = &
   - \frac{r_0^2}{2} \, \chi(t) 
   - \frac{N}{2g}
         \left  \{ 
            \chi^2(t) + \Delta D(t,t)/i
         \right \}
   \nonumber \\ &&
      + \frac{1}{2} \sum_i
         \left \{ \dot{q}_i^2(t)
               + \left . 
                    \frac{\partial^2 \, {\cal G}_{ii}(t,t')/i}
                         {\partial t \, \partial t'}
                 \right |_{t=t'}
         \right \}
   \label{eq:energyx} \\ &&
       + \frac{1}{2}  \sum_i
       \left  \{
          \chi(t) \left [ 
                      q_i^2(t) + 
                      {\cal G}_{ii}(t,t)/ i
                   \right]
           - K_{ii}(t,t,t)
       \right \}
   \>. \nonumber
\end{eqnarray}
Using the equations of motion, we can show directly that
(\ref{eq:energyx}) is conserved.  

It is easy to evaluate the energy at $t=0$ for the quantum roll
problem using the initial conditions (\ref{eq:f0dotf0}).  The result
for the leading and next-to-leading order large-N approximation is:
\[
   \frac{E}{N}
   = 
   \epsilon_0 + \frac{1}{N} \, \epsilon_1 + \cdots
\]
where,
\begin{eqnarray}
   \epsilon_0
   & = &
   \frac{1}{8G} 
   + \frac{1}{8} \, g y_0^4 
   - \frac{1}{4} \, g G y_0^2
   + \frac{1}{8} \, g G^2 \>,
   \nonumber \\
   \epsilon_1
   & = & 
   \frac{1}{4} \, g G^2 \>.
   \label{eq:eps0eps1}
\end{eqnarray}
Our initial wave function was chosen to be Gaussian, so that the
parameters of the Hartree approximation agree exactly with the energy
and parameters of the exact wave function at $t=0$.  However, the
leading order in the large-$N$ approximation for the same value of $G$
will disagree with the exact energy by $\epsilon_1$.  This discrepency
dissapears when we include the $1/N$ corrections.

Since the Hartree approximation leads to a canonical Hamiltonian
dynamical system, the corresponding energy in that approximation is
also a constant of the motion. It is given by:
\begin{eqnarray}
   E & = &
   \frac{1}{2} \,
   p_i^2
   +
   \frac{1}{8} \, G^{-1}_{ii}
   +
   2 \, \Pi_{ij} G_{jk} \Pi_{ki}
   \label{eq:EE} \\ 
   &&
   + V(q)
   + \frac{1}{2} V_{ij} G_{ij}
   \nonumber \\ &&
   + 
   \frac{1}{4!}
   V_{ijkl}
   \left (
      G_{ij} G_{kl}
      + G_{il} G_{jk}
      + G_{ik} G_{jl}
   \right ) 
   \>. \nonumber  
\end{eqnarray}
We used this expression to check the accuracy of our numerical
solutions. As with the next-to-leading order $1/N$ expression, for the
quantum roll initial condition, Eq.~(\ref{eq:EE}) agrees with the
exact result.

\section{Initial Conditions}
\label{sec:Initial}

\subsection{Quantum Roll}
\label{sec:Initialqr}

We wish to study initial conditions which are consistent with $O(N)$
symmetry.  This implies immediately that all the $x_i(t)$ have to be
identical, with $x_i(0) = 0$, and $G_{ij}(t)$ must be diagonal. The
quantum roll problem is defined by a Gaussian initial wave function
that is centered on the origin:
\begin{equation}
   \psi_0(r) 
   =
   \frac{1}{(2 \pi G)^{N/4}} \,
      \exp \left  \{ 
         - \frac{ r^2 }{ 4 G } 
           \right \} \>.
   \label{eq:psi0}
\end{equation}
In this section, $G \equiv G(0)$.   

One of the difficulties in studying the systematics of the $1/N$
expansion is the fact that, at next to leading order, every different
value of $N$ (with all other parameters held constant) defines a
different initial value problem. In this sense one cannot naively
compare individual solutions, exact or approximate, at different
values of $N$. In effect one has to tune the parameters of the problem
at each $N$ in order to maintain certain invariance properties which
allow different $N$ evolutions to be compared to each other. This
parameter tuning process is described below.

Since the infinite $N$ limit has very precise properties, several
technical issues arise when one wants to approach this limit starting
at $N=1$ in a uniform manner. To study the large $N$ limit it is
convenient to make a rescaling to the $y$ variables, given in
(\ref{eq:yscaling}). At very large $N$, the potential energy $u(y,N)$
is (\ref{eq:Uscaled}):
\begin{eqnarray}
   u(y,N) 
   & = &
   \frac{(N-1)(N-3)}{8 N^2 \, y^2} + 
   \frac{g}{8} \, ( y^2 - y_0^2 )^2  \>,
   \nonumber \\
   & \sim &
   \frac{1}{ 8\,y^2} + 
   \frac{g}{8} \, ( y^2 - y_0^2 )^2  \>.
   \label{eq:Uscaledi}
\end{eqnarray}
In this limit, $u(N,y)$ has a minimum which is independent of $N$, and
the large $N$ limit consists of harmonic oscillations about this
minimum [the reason for this is that the large $N$ limit also
corresponds to an effectively large mass limit in the Schr\"odinger
equation (\ref{eq:schroii})].

One way to uniformly study the motion of a wave packet as a function
of $N$ is to choose initial conditions so that there is a uniform
overlap of the initial wave function with the set of eigenfunctions of
the Hamiltonian in the $N \rightarrow \infty$ limit.  We can obtain
this constant overlap if we allow the coupling constant $g$ to be a
slowly varying function of $N$.  This can be done in several ways that
differ by terms of order $1/N^2$.  The method presented below leads to
uniform results even at $N=1$ as we change the parameters with $N$.
Our method is to keep the distance between the centers of the initial
wave function and the position of the minimum of the potential a
constant as $N$ is varied.

Using (\ref{eq:psi0}), we define $\tilde{r}$ by:
\begin{equation}
   \tilde{r} 
   = 
   \langle r \rangle
   =  
   \sqrt{2 G} 
   \left  [ 
      \frac{ \Gamma( (N+1)/2 ) }{ \Gamma( N/2 ) }
   \right ]  \>, 
   \label{eq:tilderi} 
\end{equation}
and $G_{\infty}$ by the variance:
\[
   \frac{G_{\infty}}{2} = 
      \langle r^2 \rangle
      - \langle r \rangle ^2 
      = G
      \left  \{ 
         N - 
         2 \left  [ 
            \frac{ \Gamma( (N+1)/2 ) }{ \Gamma( N/2 ) }
           \right ]^2  
      \right \} \>.
\]
Solving the above equations for $G$, we have
\begin{equation}
   G(N) 
   = 
   \frac{G_{\infty}}
        { 2 N - 
         4 \left  [ 
            \Gamma( (N+1)/2 ) / \Gamma( N/2 ) 
           \right ]^2  } \>.
   \label{eq:GGinfty}
\end{equation}
Substitution of this expression into (\ref{eq:tilderi}) yields
\begin{equation}
   \tilde{r}(N)
   =
   \sqrt{
   \frac{ G_\infty }
        { N \, [ \Gamma( N/2 ) / \Gamma( (N+1)/2 )]^2 - 2 } 
        } \>.
   \label{eq:tilderii} 
\end{equation}
In the limit when $N$ goes to infinity, we have
\[
   G(N)
   \rightarrow 
   G_\infty \>, \qquad
   \tilde r(N)
   \rightarrow 
   \tilde r_\infty 
   =\sqrt{(N-1) G_\infty}
   \>,
\]
which defines $\tilde r_\infty$, and agrees with the asymptotic form
of the rescaled version of the initial wave function (\ref{eq:psi0}):
\begin{eqnarray*}
   \phi_0(r) 
   &=& 
   \frac{1}{(2 \pi G)^{N/4}} \,
      \exp \left  \{ 
         - \frac{ r^2}{ 4 G} + \frac{N-1}{2} \ln r  
           \right \}  \>, \\
   & \approx & 
   \frac{1}{(2 \pi G)^{N/4}} \,
      \exp \left  \{ 
         - \frac{ (r - \tilde{r}_{\infty})^2 }{ 2 G_\infty } 
         + O(1/\sqrt{N} )
           \right \}  \>.
\end{eqnarray*}
In order to ensure that the initial wave function has a finite overlap
with the energy eigenfunctions of the Schr\"odinger equation at large
$N$, we will keep the value of $G_\infty$ (and not $G$) fixed in our
simulations.

Another quantity that should be kept constant is the basic oscillation
frequency. In order to do this, we first find the Gaussian
oscillations about the minimum of the one dimensional potential,
defined by Eq.  (\ref{eq:Uofr}).  We expand $U(r)$ as
\begin{equation}
   U(r) = U(\bar{r}) +
      \frac{1}{2} \bar{m}^2 \, ( r - \bar{r} )^2 + \cdots
   \>,
   \label{eq:Uofr_exp}
\end{equation}
where $\bar{r}$ is given by the solution of the equation,
\begin{equation}
   \frac{(N-1)(N-3)}{4 \, \bar{r}^4} 
   = \frac{g}{2 \, N} \, ( \bar{r}^2 - r_0^2 )  \>,
   \label{eq:barrmin}
\end{equation}
and $\bar{m}^2$ by
\begin{eqnarray}
   \bar{m}^2 
   &=& \frac{3 \, (N-1)(N-3)}{4 \, \bar{r}^4 } + 
      \frac{g}{2 \, N} \, (3 \, \bar{r}^2 - r_0^2 ) \>, 
   \label{eq:barmr0} \\
   &=& \frac{g}{N} \, ( 3 \, \bar{r}^2 - 2 \, r_0^2 )  \>.
   \nonumber 
\end{eqnarray}
The frequency of oscillation is determined by $\bar{m}$, and this is
the quantity to be kept fixed as $N$ is changed.

The last technical issue is to keep the distance between the center of
the initial wave function, $\tilde r$, and the minimum of the
potential, $\bar r$, a constant as we vary~$N$.  That is, we keep
\[
   \delta r = \bar{r} - \tilde{r} \>,
\]
constant for all $N$.  

With this strategy of keeping $G_\infty$, $\bar{m}$ and $\delta r$
fixed, we can now determine how the coupling constant must vary 
with~$N$.  We first define $m^2$ to be the second derivative of $V(r)$
evaluated at $r = \bar{r}$,
\begin{equation}
   m^2 
   = 
   \frac{{\rm d}^2 V(r)}{{\rm d} r^2} 
   \bigg |_{r = \bar{r}} 
   = 
   \frac{g}{2 N} \, ( 3 \, \bar{r}^2 - r_0^2 ) \>.
   \label{eq:mtr0}
\end{equation}
Then, (\ref{eq:barmr0}) becomes
\begin{equation}
   m^2(N) = \bar{m}^2 - \frac{ 3 \, (N-1)(N-3) }{ 4 \, \bar{r}^4 }  \>.
   \label{eq:mtwo}
\end{equation}
Solving (\ref{eq:mtr0}) for $r_0$, substituting into
(\ref{eq:barrmin}) and solving for $g$ gives:
\begin{equation}
   g(N) 
   =  
   \frac{N}{ \bar{r}^2 } 
      \left  \{ 
         \bar{m}^2 - \frac{ (N-1)(N-3) }{ \bar{r}^4 } 
      \right \}  \>.
   \label{eq:grunning}
\end{equation}
with $\bar{r} = \tilde{r} + \delta r$.  The value of $r_0^2$ is then
determined by (\ref{eq:barrmin}):
\begin{equation}
   r_0^2(N) 
   = 
   \bar{r}^2 - 
   \frac{2 N}{g(N)} \, \frac{(N-1)(N-3)}{4\, \bar{r}^4}  \>.
   \label{eq:r02}
\end{equation}
Thus, for {\em fixed} values of $G_\infty$, $\bar{m}$ and $\delta r$,
Eqs.~(\ref{eq:GGinfty}), (\ref{eq:tilderii}), (\ref{eq:grunning}), and
(\ref{eq:r02}) determine values for $G(N)$, $\tilde{r}(N)$, $g(N)$,
and $r_0(N)$ all values of $N$.

In the limit, $N \rightarrow \infty$, we find that
\begin{eqnarray}
   g(\infty) 
   & = & 
   \frac{1}{G_\infty} 
   \left  (
      \bar{m}^2 - \frac{1}{G_\infty^2} 
   \right )
   \label{eq:ginfty} \\
   m^2(\infty) 
   & = & 
   \bar{m}^2 - \frac{3}{4 \, G_\infty^2}
   \label{eq:m2infty}
\end{eqnarray}

\begin{figure}
   \centering
   \epsfig{figure=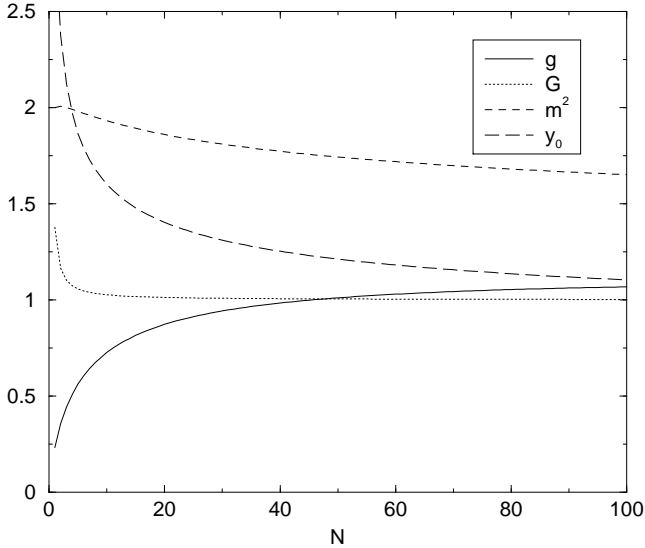,width=3.35in}
   \caption{Potential parameters as a function of $N$.}
   \label{fig:potparams}
\end{figure}

To summarize, in order to establish appropriate initial conditions for the
quantum roll problem, we have kept the variance $G_{\infty}$ constant
instead of $G$, and have allowed the parameters describing the
potential function $g$ and $r_0$ to change with $N$ in order to
compare solutions that have close to the same oscillation frequencies.
In our numerical runs, we chose the values
\begin{equation}
G_\infty = 1;~~ \bar{m}^2 = 2;~~\delta r =
2;~~g(\infty) = 1  \label{eq:values}
\end{equation}
Fig.~\ref{fig:potparams} displays the variation of the potential
parameters with $N$.

\subsection{Shifted Gaussian Initial Conditions}
\label{sec:Initialsg}

The second $O(N)$ invariant initial condition we investigated had a
wave function localized in a wave packet near the center of the valley
of the classical potential at $r = r_0$.  For $N=1$ this would be the
standard double-well tunnelling problem; for higher values of $N$,
tunneling is avoided by going around the barrier. Therefore this
initial condition is qualitatively different from the roll problem and
provides a different arena for testing approximations. However, since
this initial condition violates the $O(N)$ symmetry of the potential,
numerical solution is at present possible only for very small $N$.

We take the initial wave function to be a shifted Gaussian of the
form:
\begin{equation}
   \psi_0(x) = 
   \frac{1}{(2 \pi G)^{N/4}} \,
   \exp \left  \{
      - \sum_i \frac{ (x_i - r_0/\sqrt{N})^2 }{ 4 G }
        \right \} 
   \>.
   \label{eq:psi0sg}
\end{equation}
The energy $E$ of this state can be determined from Eq.~(\ref{eq:EE})
by the substitutions,
\[ 
   G_{ij} \rightarrow \delta_{ij} G \>; \quad
   q_{i}   \rightarrow  r_0/\sqrt{N} \>; \quad
   p_{i}   \rightarrow 0 \>; \quad
   \Pi_{ij} \rightarrow  0  \>,
\]
from which we find:
\begin{equation}
   E
   = 
   \frac{N}{8G} + 
   \frac{g}{8N}
      \biggl\{
         N (N + 2) \, G^2 - 2 N \, G \, r_0^2 + r_0^4
      \biggr\}
   \>.
\label{eq:E_ini}   
\end{equation}
On the other hand, the height of the classical potential barrier is
given by:
\begin{equation}
   E_b =  \frac{g}{8 N} \, r_0^4 \>.
   \label{eq:E_bar}   
\end{equation}

For $N=1$, the necessary requirement for tunneling is that $E < E_b$. 

In the general case (arbitrary $N$), we have:
\begin{equation}
   M^2 =
 \frac{\partial^2 V(r)}{\partial r^2} \bigg |_{r_0}
   =
   \frac{g}{N} \, r_0^2 
   \qquad {\rm or} \qquad
   r_0^2 = \frac{N}{g} \, M^2 
   \>.
\label{eq:M2_def}   
\end{equation}
If the initial state is close to the ground state of a harmonic
potential that approximates the potential at the bottom of the well,
then the width $G$ of the wave function is, approximately,
\begin{equation}
   G = \frac{1}{2 \sqrt{M^2}} \>,
   \label{eq:G0_def}   
\end{equation}
which can be combined with Eq.~(\ref{eq:E_ini}) to give the desired
energy of the initial state in terms of the values of $N$ and $g$.

We are interested in initial conditions where the energy per
oscillator does not increase as a function of $N$.  To implement this
we fix $M^2 = 1$,  which corresponds to $G = 1/2$ for the 
initial width.  The barrier height is then given by
\begin{equation} 
   E_b = \frac{N}{8g}  \>, 
   \label{eq:EbNg}
\end{equation}
and the total energy by
\begin{equation} 
   E = \frac{N+1}{4} + \frac{N+2}{32} g 
   \label{eq:ENg}
\end{equation}
We explored three cases: $E = 0.5 \, E_b$, $E = E_b$, and $ E = 2 \,
E_b$.  For each of these cases, Eqs.~(\ref{eq:EbNg}) and
(\ref{eq:ENg}) determine $g$ for each $N$.  In all cases we took
$x_i(0) = r_0 / \sqrt{N}$, $\dot{x}_i = 0$, $G_{ij}(0) = G
\delta_{ij}$, and $\dot{G}_{ij}(0) = 0$.  As a consequence, all of the
oscillators $x_i(t)$ move identically.

\section{Effective potential}
\label{sec:effpot}

It is well-known that the static effective potential is not always a
useful guide to the true dynamics of the system (See, e.g.,
Ref.~\cite{ref:CHKM}). Nevertheless, one may seek to gain qualitative
insight into some aspects of quantum dynamics this way, though care is
certainly indicated (See, e.g., Ref.~\cite{ref:S} for the Gaussian
effective potential). Indeed, there appears to be an interesting
connection with the properties of the effective potential at
next-to-leading order and with the corresponding dynamical evolution
(discussed in the next section).

The effective potential in the large $N$ approximation has been
previously obtained by Root \cite{ref:ROOT} to order $1/N$; however,
we recalculate it here using our equations.  When $x_i$ and $\chi$ are
independent of time, we can ignore the closed time path ordering and
use Fourier transforms, passing the poles by using the Feynman
contour.  Then, from the action given in (\ref{eq:effaction}), we
find
\begin{eqnarray}
   \lefteqn{
   V_{\text{eff}}^{[1]}(r,\chi)  =
   \frac{N \chi}{g} \,
   \left ( \mu^2 - \frac{\chi}{2} \right )
   + 
   \frac{1}{2} \, \chi \, r^2 }
   \label{eq:Veff} \\ 
   &&
   +
   \frac{N}{2} \int \frac{dk}{2\pi i} \, \ln [ \tilde G^{-1}(k) ]
   + 
   \frac{1}{2} \int \frac{dk}{2\pi i} \, \ln [ \tilde D^{-1}(k) ]
   \>,
   \nonumber
\end{eqnarray}
where $\chi$ satisfies the requirement
\begin{eqnarray}
   \frac{\partial}{\partial \chi} V_{\text{eff}}(r,\chi)
   = 
   0 \>.
   \label{eq:pVeffpChi}
\end{eqnarray}
In this section to make contact with Root \cite{ref:ROOT}, we have
$\mu^2 = - g r_0^2/(2N) < 0$.  In order to examine the large $N$
limit, we again rescale (\ref{eq:yscaling}) to the $y$ variables.
Then for the Green's functions, we find:
\begin{eqnarray*}
   \tilde{G}^{-1}(k) 
   & = &
   - (k^2 - \chi)
   \>,
   \\
   \tilde{D}^{-1}(k) 
   & = &
      - \frac{N}{g} - N y^2 \, \tilde{G}(k)
      + \frac{i N}{2} \,
      \int \ \frac{dp}{2 \pi} \ \tilde{G}(p) \, \tilde{G}(k-p)
   \\ 
   & = &
   - \frac{N}{g} 
   \Biggl \{ 
      1 - 
      g \, \frac{y^2}{k^2 - \chi} - 
      \frac{g}{2 \sqrt{\chi}} \, \frac{1}{k^2 - 4 \chi}
   \Biggr \}
   \\ 
   & = &
   - \frac{N}{g}
   \frac{(k^2 - m_{+}^{2}) (k^2 - m_{-}^{2})}
        {(k^2 - 4 \chi) (k^2 - \chi)} \>, 
\end{eqnarray*}
where $m_{\pm}^2 = b \pm \sqrt{ b^2 - c }$, with
\begin{eqnarray*}
   b
   & = & 
   \frac{5}{2} \, \chi + 
   \frac{g}{2} \left ( y^2 + \frac{1}{2\sqrt{\chi}} \right )
   \\
   c
   & = & 
   4 \, \chi^2 + 
   g \left ( 4 \, y^2 \chi + \frac{1}{2} \sqrt{\chi} \right ) \>.
\end{eqnarray*}
For the Feynman contour, we have
\begin{equation}
   \int \ \frac{dk}{2\pi i} \ \ln (k^2 - \chi)
   = 
   \sqrt{\chi} + \text{constant terms}
   \>.
\end{equation}
Thus the effective potential (\ref{eq:Veff}) becomes
\begin{eqnarray}
   {V_{\text{eff}}^{[1]}(y,\chi)  \over N} 
   & = &
   \frac{\chi}{2} \, \bigl ( y^2 - y_0^2 \bigr ) -
   \frac{\chi^2}{2 \, g} + 
   \frac{\sqrt{\chi}}{2}
   \label{eq:Veff_sec} \\
   && \qquad {}+
   \frac{1}{2 \, N} \, 
      \bigl ( m_{+} + m_{-} - 3 \, \sqrt{\chi} \, \bigr )  \>.
   \nonumber 
\end{eqnarray}
The gap equation which determines $\chi$ follows from
(\ref{eq:pVeffpChi})
\begin{eqnarray}
   \chi 
   & = &
   \frac{g}{2} \, \bigl ( y^2 - y_0^2 \bigr ) +
   \frac{g(N - 3)}{4 \, N \,\sqrt{\chi} } + 
   \frac{g}{2\,N} \, 
      \frac{\partial (m_{+} + m_{-})}{\partial \chi} \>.
   \label{eq:chi_sec}
\end{eqnarray}
To leading order in the large $N$ expansion, Eqs.~(\ref{eq:Veff_sec},
\ref{eq:chi_sec}) reduce to the parametric set
\begin{eqnarray}
   {V_{\text{eff}}^{[0]}(\chi) \over N}
   & = &
   \frac{\chi^2}{2g} + \frac{ \sqrt{\chi}}{4} \>, 
   \nonumber \\
   y^2(\chi)
   & = &
   y_0^2 + \frac{2}{g} \, \chi - \frac{1}{2 \, \sqrt{\chi}}
   \>. \label{eq:Vchi0}
\end{eqnarray}
Equations (\ref{eq:Veff_sec}) and (\ref{eq:chi_sec}) agree with Root,
however he used the leading order expression for $\chi$ in
(\ref{eq:Vchi0}), rather than the full~$\chi$ of (\ref{eq:chi_sec}).

There exist two real solutions of Eq.~(\ref{eq:chi_sec}) for $\chi$
with $y$ greater than some minimum value $y_{\text{min}}$.  The
next-to-leading order large $N$ effective potential, from
Eq.~(\ref{eq:Veff_sec}) is therefore double valued for $y >
y_{\text{min}}$, and does not exist for smaller values of $y$. The
physical solution branch corresponds to the one that matches on to the
leading order result; the other branch is an unphysical
solution. Since it follows from a Legendre transformation, the
effective potential (at any order in $1/N$) has to be a convex
function. The nonexistence of the effective potential at $y <
y_{\text{min}}$ implies that no quantum state can be associated with
the next-to-leading order large $N$ approximation in this range.

\begin{figure}
   \centering \epsfig{figure=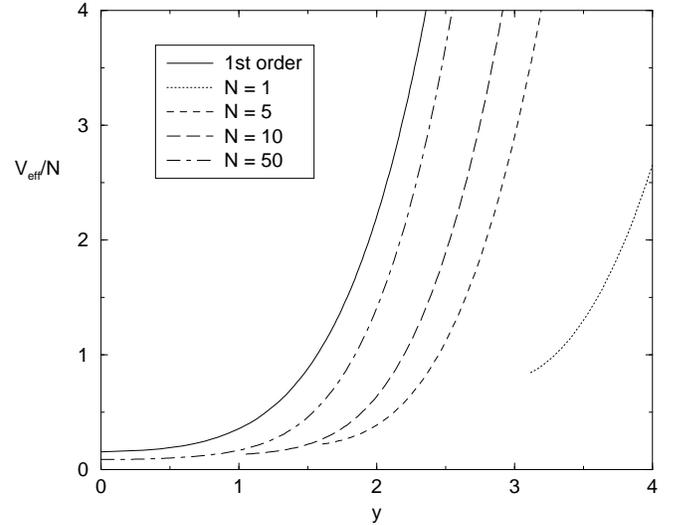,width=3.35in}
   \caption{$V_{\text{eff}}/N$ vs $y = r / \sqrt{N}$ for the leading
   and next-to\-leading order large-$N$ approximation for different
   values of $N$.}  
\label{fig:effpot_largeN}
\end{figure}

In Fig.~\ref{fig:effpot_largeN}, we plot the physical branch of the
effective potential as a function of $y$, for values of $N$ from $1$
to $100$, for the case $g = 1$ and $y_0 = 2$. For comparison, we also
show in this figure the leading order potential function from
Eq.~(\ref{eq:Vchi0}), which is single-valued and finite for all
$y$. (In contrast to the next-to-leading order case we can always
associate a Gaussian wave function with the leading order
approximation.)

In the case of the Hartree approximation, one can define an
``effective potential'' as the expectation value of the Hamiltonian
using the variational wave function (\ref{eq:trial}) for static
configurations \cite{ref:S,ref:sat}.  Setting $p_i(t) = 0$ and
$\Pi_{ij}(t) = 0$, and putting $\sum_i q_i^2 = r^2$ and $G_{ij} =
\delta_{ij} G$ in Eqs.~(\ref{eq:kin_erg}) and (\ref{eq:expcVHartree}),
we find:
\begin{eqnarray}
   \frac{V_{\text{eff}}^{[H]}(y,G)}{N} 
   & = & 
   \frac{1}{8G} + \frac{g}{8} \left ( y^2 - y_0^2 \right )^2
   \label{eq:VHeff_0}  \\
   && 
     {}+  g \frac{N+2}{4 N} 
         \left  ( 
            y^2 + \frac{1}{2} \, G 
         \right ) \, G
      -  \frac{g}{4} \, y_0^2 \, G \>,
   \nonumber
\end{eqnarray}
The value of $G$ is fixed by the requirement that
\begin{displaymath}
   \frac{ \partial \, V_{\text{eff}}^{[H]}(y,G)}
        { \partial G} = 0 \>,
\end{displaymath}
which gives the gap equation for the Hartree approximation:
\begin{equation}
   \chi 
   =
   \frac{g}{2} ( y^2 - y_0^2 ) + 
   \frac{g}{N} \left ( y^2 + \frac{1}{2\chi} \right ) + 
   \frac{g}{4 \sqrt{\chi}} \>,
   \label{eq:Hgap}
\end{equation}
where we have set $G = 1/ 2 \sqrt{\chi}$.  Parametric equations for
the Hartree effective potential are then given by:
\begin{eqnarray}
   \frac{V_{\text{eff}}^{[H]}(\chi)}{N} 
   & = & 
   \frac{1}{2g}\,\left ( \frac{N}{N + 2} \right )^2 \, \chi^2 - 
   \frac{N}{(N + 2)^2} \, y_0^2 \, \chi 
   \nonumber \\ 
   &&{}+ 
   \frac{N + 4}{4 (N + 2) } \, \sqrt{\chi} + 
   \frac{g}{2 ( N + 2 )^2} \, y_0^4 
   \nonumber \\ 
   &&{}+ 
   \frac{g}{4 ( N + 2 )} \, \frac{y_0^2}{\sqrt{\chi}} - 
   \frac{g}{16 \, N} \, \frac{1}{\chi}
   \label{eq:VeffHpara} \\
   y^2(\chi)
   & = & 
   \frac{N}{N+2} \, 
      \left ( 
         y_0^2 + \
         \frac{2}{g} \, \chi -
         \frac{1}{2 \, \sqrt{\chi}}
      \right ) - 
   \frac{1}{(N + 2) \, \chi} \>.
   \nonumber 
\end{eqnarray}
Note that in the limit $N \rightarrow \infty$,
Eq.~(\ref{eq:VeffHpara}) reduces to Eq.~(\ref{eq:Vchi0}), the leading
order large $N$ result. Note also that the Hartree effective potential
is not derived from a Legendre transform and hence is not subject to a
convexity constraint.

\begin{figure}
   \centering
   \epsfig{figure=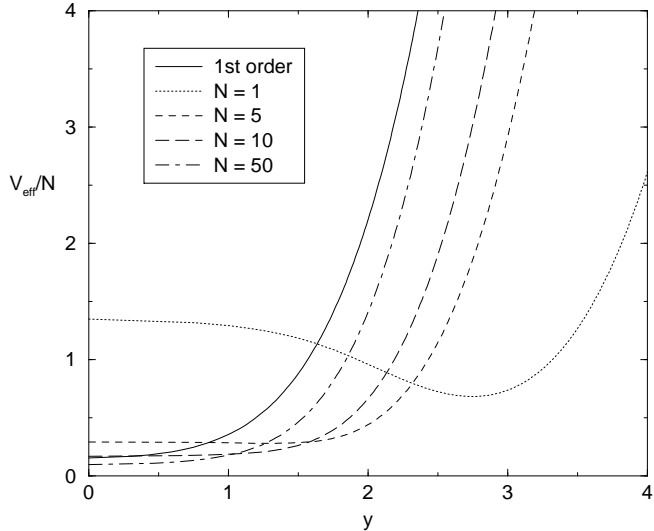,width=3.35in}
   \caption{$V_{\text{eff}}/N$ vs $y = r / \sqrt{N}$ 
   for the Hartree approximation using the parameters found in Eq.
(\ref{eq:values}).}    \label{fig:effpot_Hartree}
\end{figure}

In Fig.~\ref{fig:effpot_Hartree}, we plot the Hartree effective
potential from Eq.~(\ref{eq:VeffHpara}) as a function of $y$, for our
chosen parameters $g_{\infty} = 1$ and $y_0 = 2$, for different values
of $N$. In contrast to the smooth behavior exhibited by the large $N$
effective potentials, the Hartree effective potential shows a
``first-order transition'' in the placement of the minimum of the
potential as a function of $N$.

\begin{figure}
   \centering
   \epsfig{figure=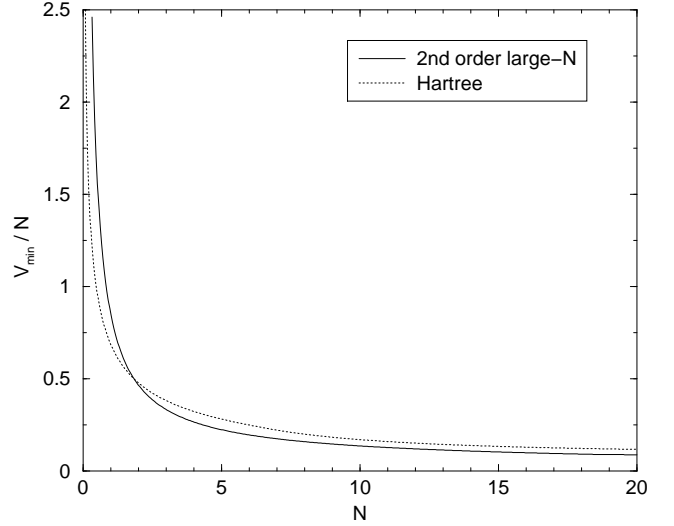,width=3.35in}
   \caption{$V_{\text{eff}} / N$ at the minimum for the Hartree  
     and 2nd order large-$N$ effective potentials, as a
     function of $N$, for the same set of parameters as in Fig.
\ref{fig:effpot_Hartree}.}    \label{fig:effpot_energies}
\end{figure}

The minimum of the effective potential corresponds to a determination
of the ground state energy. In Fig.~\ref{fig:effpot_energies}, we show
values of the minimum energies of the large $N$ and Hartree effective
potentials as a function of $N$. For $N \ge 2$, the Hartree minimum is
generally greater than that for the next-to-leading order large-$N$
approximation. Since the Hartree approximation is a variational ansatz
it gives an upper bound to the minimum energy. The fact that the
next-to-leading order large $N$ results are lower than this bound is
encouraging, although no guarantee of absolute accuracy.

\begin{figure}
   \centering
   \epsfig{figure=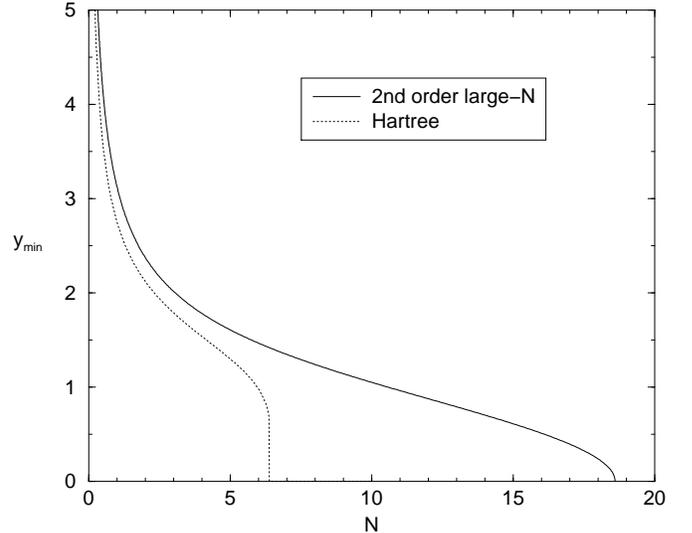,width=3.35in}
   \caption{The value of $y$ at the minimum of the effective
     potential for the Hartree and 2nd order large-$N$
     approximation, as a function of $N$, for the
same set of parameters as in Fig.~\ref{fig:effpot_Hartree}.
     }
   \label{fig:effpot_rmin}
\end{figure}

It is interesting to ask the question how the point $y_{\text{min}}$,
below which the next-to-leading order large $N$ effective potential
does not exist, changes as a function of $N$. We know that at
``infinite $N$,'' (leading order), $y_{\text{min}}=0$ but it is
important to know how this limit is reached. For instance, is there a
finite value of $N$ beyond which $y_{\text{min}}=0$? In
Fig.~\ref{fig:effpot_rmin}, we plot $y_{\text{min}}$ as a function of
$N$, for the Hartree and next-to-leading order large $N$
approximations. As already stated, the Hartree approximation displays
a first order phase transition between the broken and unbroken
symmetry solutions at $N = 6.2$, whereas for the next-to-leading order
large $N$ approximation a different type of behavior is found: for $N
\le 18.6$, $y_{\text{min}}$ is finite, but for $N \ge 18.6$, it hits
the origin. Thus for $N \ge 18.6$, we can associate a quantum state
(though not known explicitly) with the next-to-leading order
approximation.

The critical value of $N$ is fixed by the value of $\chi$ at the gap
equation at the inflection point.  If we write the gap equation
(\ref{eq:chi_sec}) as
\begin{equation}
   f(\chi, y^2, N) 
   = f_0(\chi, y^2) + \frac{1}{N} \, f_1(\chi, y^2)
   = 0 \>,
\end{equation}
where
\begin{eqnarray*}
   f_0(\chi, y^2)
   & = &
   \frac{g}{2} \, ( y^2 - y_0^2 ) + 
   \frac{g}{4} \, \frac{1}{\sqrt{\chi}} - \chi  \>,
   \\
   f_1(\chi,y^2)
   & = &
   - \frac{3\, g}{4} \, \frac{1}{\sqrt{\chi}} + 
   \frac{g}{2} \, \frac{\partial ( m_{+} + m_{-} ) }{\partial \chi}
   \>.
\end{eqnarray*}
Then the critical point is determined by
\begin{equation}
   N_{\text{c}} 
   = 
   - \frac{ f_1(\bar\chi,0)}
          { f_0(\bar\chi,0)}
   =
   - \frac{   \partial f_1(\bar\chi,0) / 
              \partial \bar\chi   }
          {   \partial f_0(\bar\chi,0) /
              \partial \bar\chi   }  \>,
   \label{eq:Ncrit}
\end{equation}
where $\bar{\chi}$ is given by the solution of this system of
equations.  For the parameters of Eq. (\ref{eq:values}), we find
numerically that $N_{\text{c}} = 18.60$ in excellent agreement with
the results shown in Fig.~\ref{fig:effpot_rmin}.

\section{Numerical Results}
\label{sec:Res}

\subsection{Quantum Roll}

\begin{figure}
   \centering {\epsfig{figure=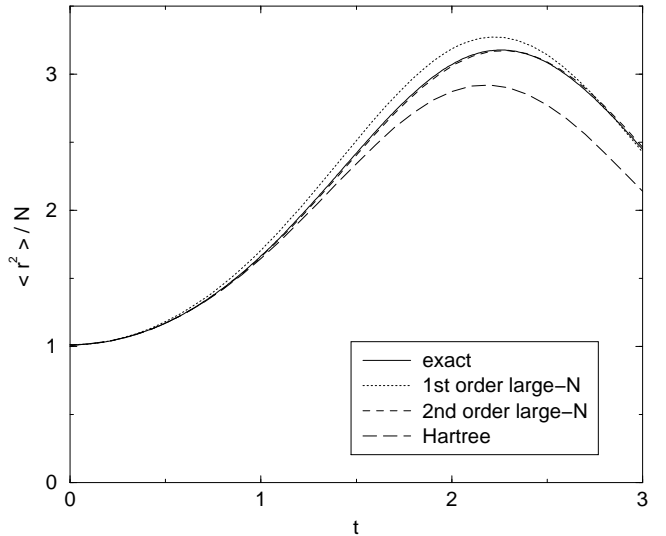,width=3.35in}}
   \caption{$\langle r^2 \rangle / N$ for $N=20$, for the exact,
   leading and next-to-leading order large-$N$, and Hartree
   approximations for short times.}  \label{fig:raw_short}
\end{figure}

We begin with a discussion of our results for the quantum roll
problem. We first examine the short time behavior, $0 < t < 3$, to see
if the next-to-leading order large $N$ approximation gives an
improvement over the leading order solutions.  In
Fig.~\ref{fig:raw_short}, we plot the values of $\langle r^2 \rangle /
N$ from the numerical solution, the leading and next-to-leading order
large-$N$ approximations, and the Hartree approximation, for $N =
20$. The next-to-leading order large $N$ approximation is clearly
better than the leading order solution, and also better than the
Hartree results. Similar behavior is seen for other values of $N$ (we
also ran $N=50,~80,$ and 100).

\begin{figure}
   \centering
   {\epsfig{figure=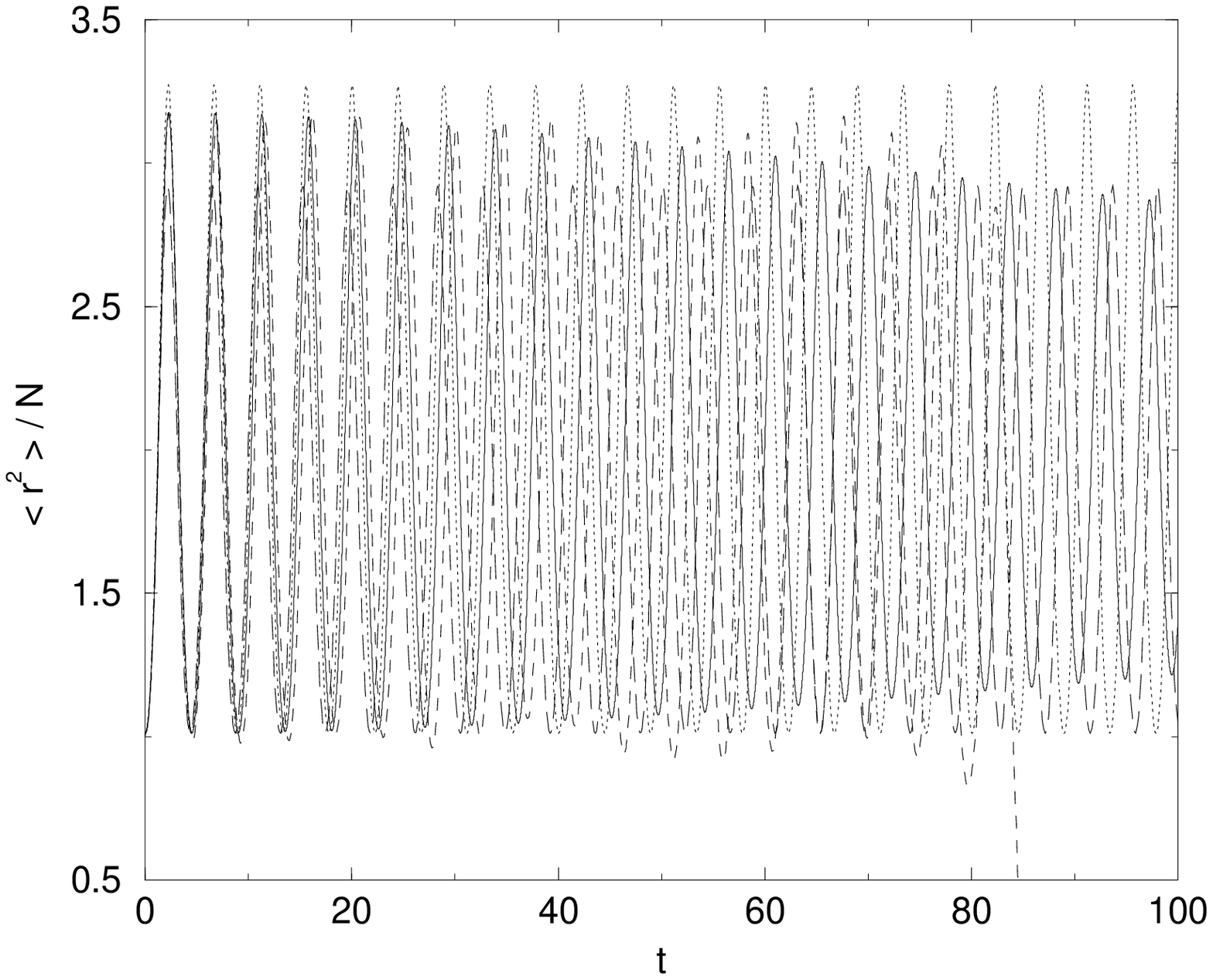,width=3.35in,height=2.0in}}
   {\epsfig{figure=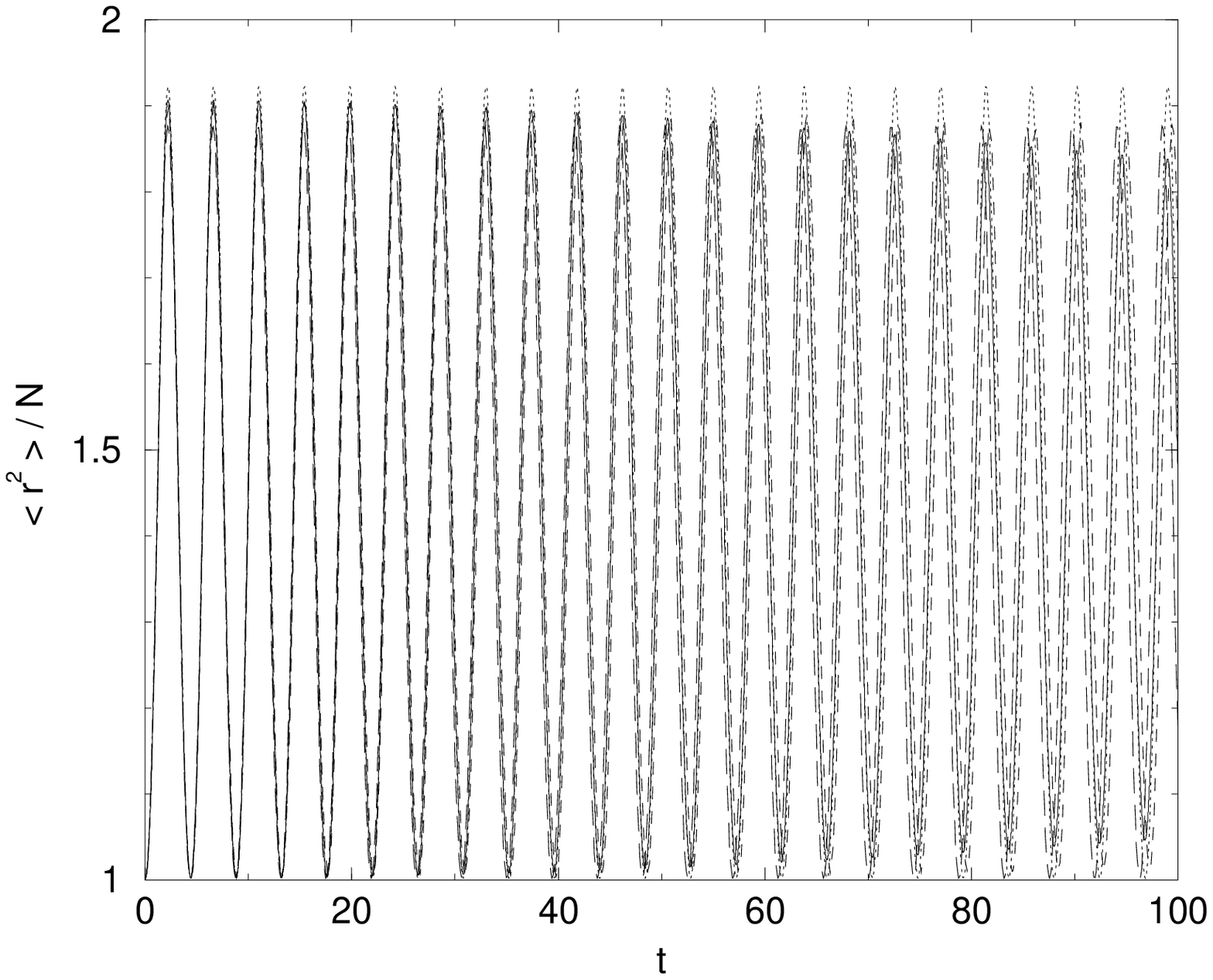,width=3.35in,height=2.0in}}
   \caption{$\langle r^2 \rangle / N$ for the exact, leading and
   next-to-leading order large-$N$, and Hartree approximations for
   long times. The top figure is for $N=20$ while the bottom figure is
   for $N=100$. The labeling conventions are the same as in the
   previous figure.}  \label{fig:raw_long}
\end{figure}

The long time behavior of these approximations is typically of much
more interest. We examined behavior over times $0 < t < 100$ to see
how long the approximations remained viable.  Fig.~\ref{fig:raw_long}
displays $\langle r^2 \rangle / N$ for the numerical solution, the
leading and next-to-leading order large-$N$ approximations, and the
Hartree approximation for $N = 20$ and 100. The next-to-leading order
large-$N$ approximation for $N = 20$ blows up at $t \sim 84$. This
instability is connected to a violation of unitarity in the particular
implementation of this approximation and will be discussed in greater
detail below.  In general, at these moderate values of $N$, the
approximations track the numerical solutions reasonably well though
they do get out of phase as time progresses. As $N$ is increased, the
phase errors are considerably reduced as is apparent in the results
for $N=100$.

The energy of the next-to-leading order large $N$ and Hartree
approximations is the same as the exact one, but the energy of the
leading order large $N$ approximation differs from it by terms of
order $1/N$. (This is because we need to keep the initial values of
the parameters the same.) To make a comparison between the
approximations this difference has to be compensated for; we do this
by rescaling time by a constant multiplicative factor so as to match
the last oscillation maxima.  This effect is of order $1/N$. For
$N=100$, we find that for $0 < t < 100$, the next-to-leading order
$1/N$ approximation is always more accurate than the leading order;
however, when comparing the next-to-leading order with the Hartree,
although less accurate for $t < 50$, the Hartree approximation starts
becoming more accurate at $t \sim 50$ (however, the errors are very
similar in magnitude, of the order of a few per cent).

\begin{figure}
   \centering \epsfig{figure=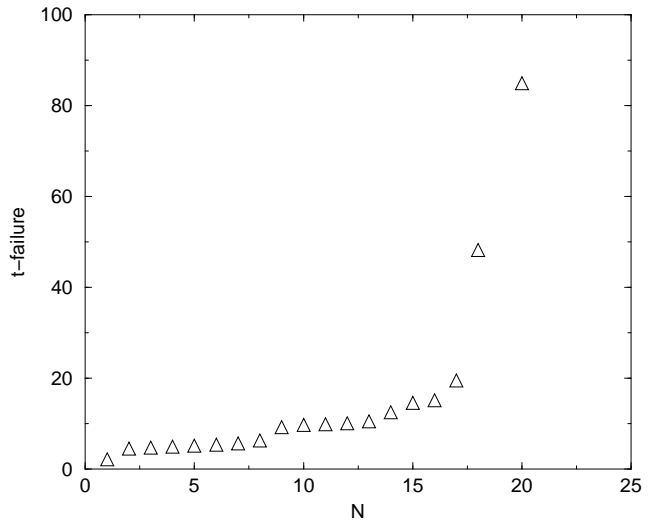,width=3.35in}
   \caption{Failure time for the next-to-leading order large $N$
   approximation, as a function of $N$, for the same set of parameters
   as in Fig. \ref{fig:effpot_Hartree}.  Here failure time is defined
   as the time at which $\langle r^2 \rangle / N$ becomes negative.}
   \label{fig:tfailure}
\end{figure}

We now return to a discussion of the blow ups first encountered in the
$N=20$ case discussed above. The failure of truncation schemes (of
which $1/N$ is an example) to maintain a rigid connection with the
existence of a probability distribution function is a well-known
problem in nonequilibrium statistical mechanics. It is often the case
that, when this connection is lost (failure of reality or positivity
conditions), instability soon follows. In our case, the violation of
unitarity is manifest in that positive expressions such as $\langle
r^2 \rangle / N$ can turn out to be negative. Note that since both the
Hartree and leading order $1/N$ approximations are variational in
nature, they can never violate unitarity. In Fig.~\ref{fig:tfailure},
we show the blow up or failure time for the next-to-leading order
large $N$ approximation as a function of $N$; the failure time being
defined as the time at which $\langle r^2 \rangle / N$ becomes
negative.  It is interesting to note that for values of $N$ near the
critical value $N_{\text{c}} = 18.60$ where the effective potential
extends down to the origin, the failure time starts to increase
rapidly (consistent with our interpretation of the connection of the
static effective potential to an associated quantum state).  For
values of $N$ greater than 21, we could not find failure for $t <
150$. Thus, it may well be that above a certain value of $N$, the
unitarity violation is pushed out to times of no practical
significance, or may even disappear altogether.

\begin{figure}
   \centering {\epsfig{figure=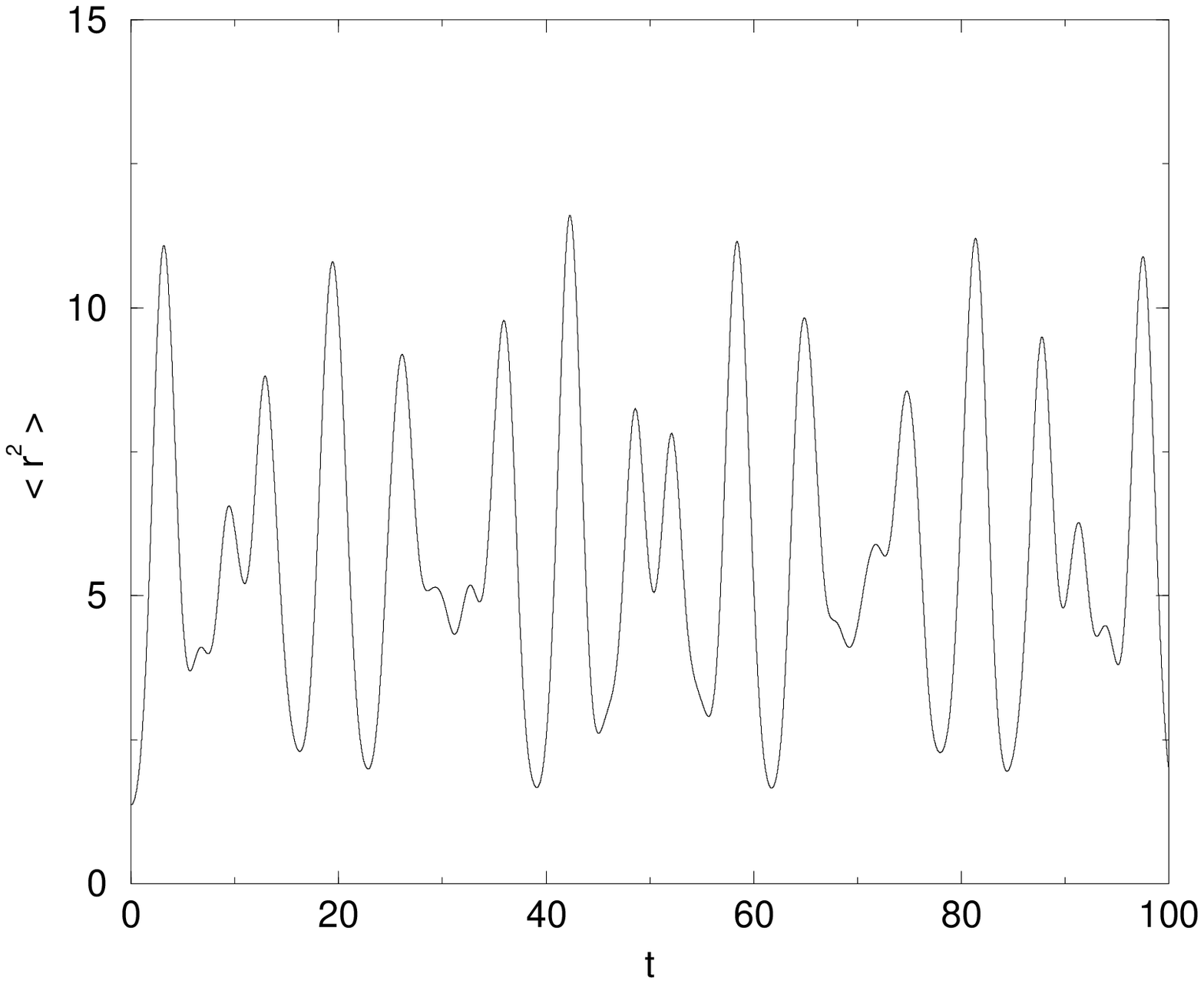,width=3.35in,height=1.5in}}
   {\epsfig{figure=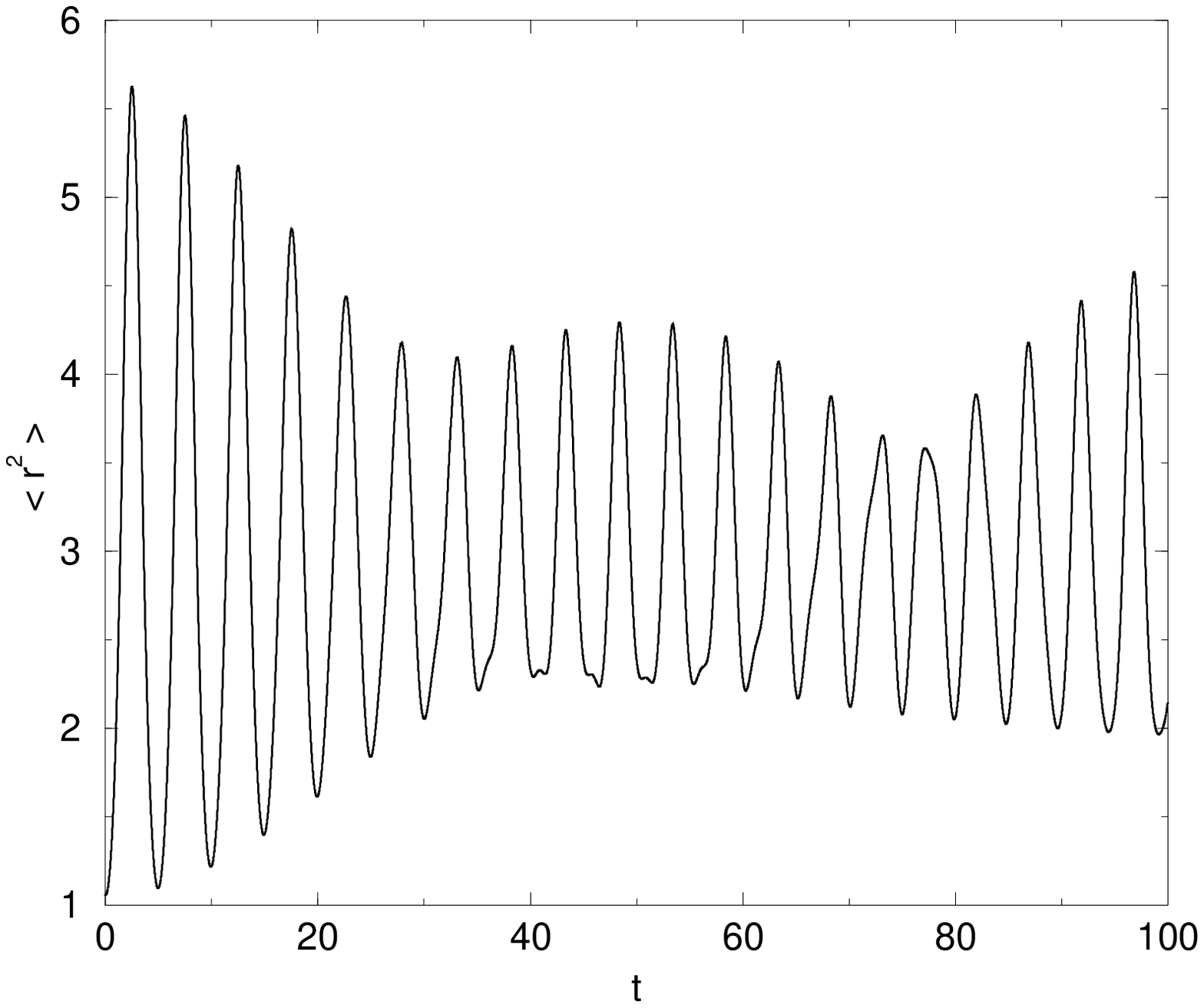,width=3.35in,height=1.5in}}
   {\epsfig{figure=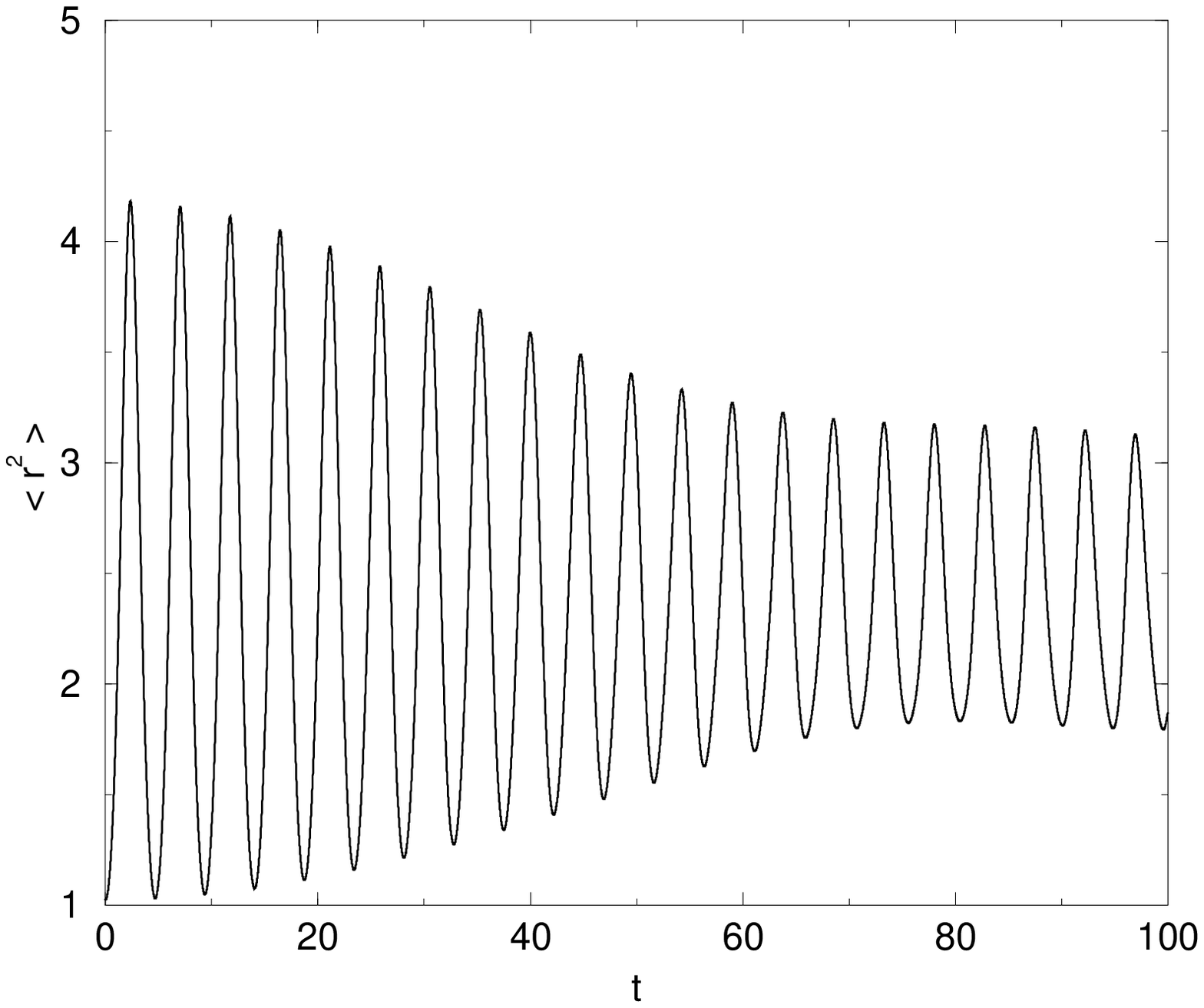,width=3.35in,height=1.5in}}
   {\epsfig{figure=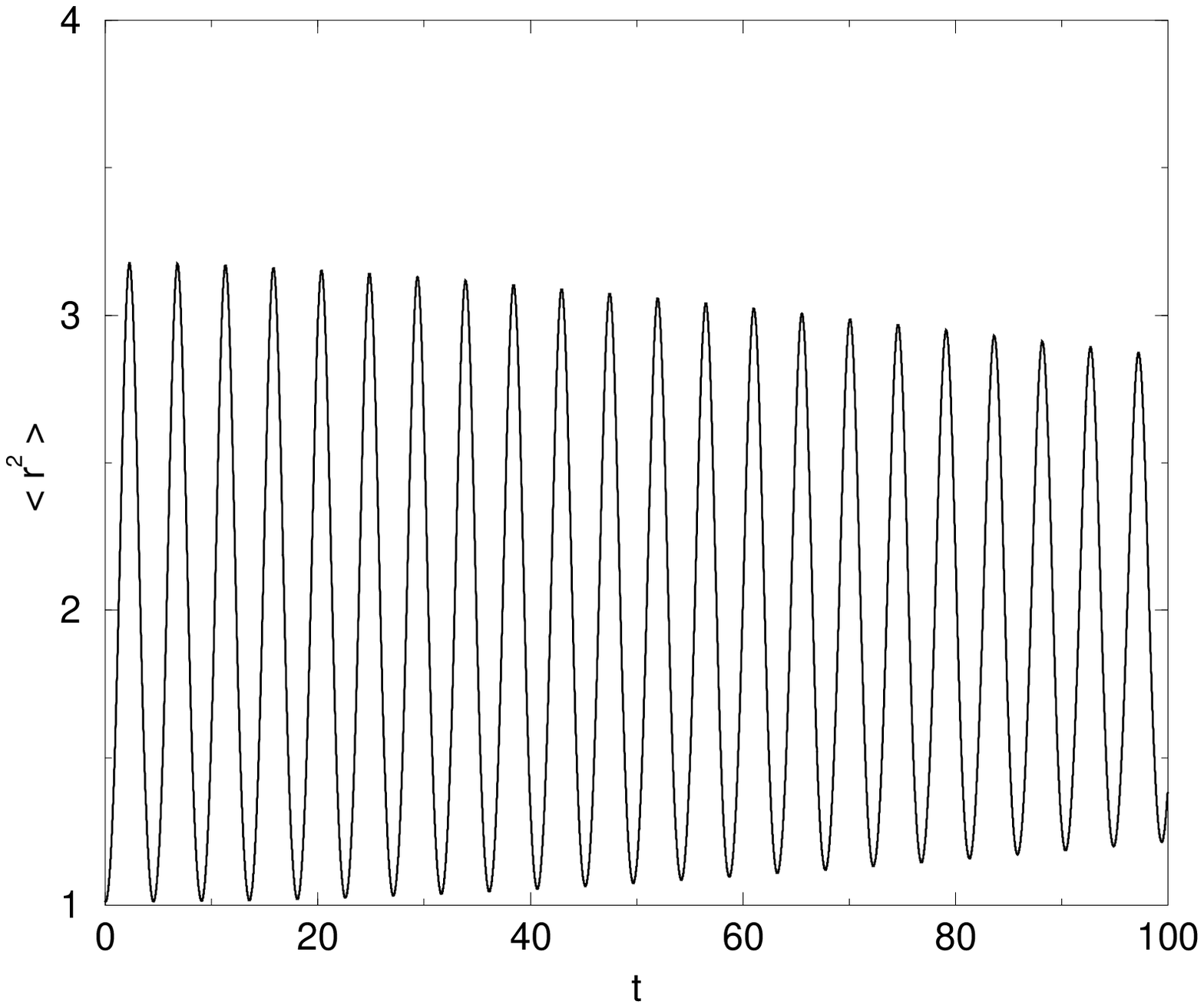,width=3.35in,height=1.5in}}
   \caption{Exact solutions for $\langle r^2 \rangle / N$ as a
   function of time. From top to bottom, $N=1,~5,~10,$ and 20.}
   \label{fig:exact}
\end{figure}

\begin{figure}
   \centering {\epsfig{figure=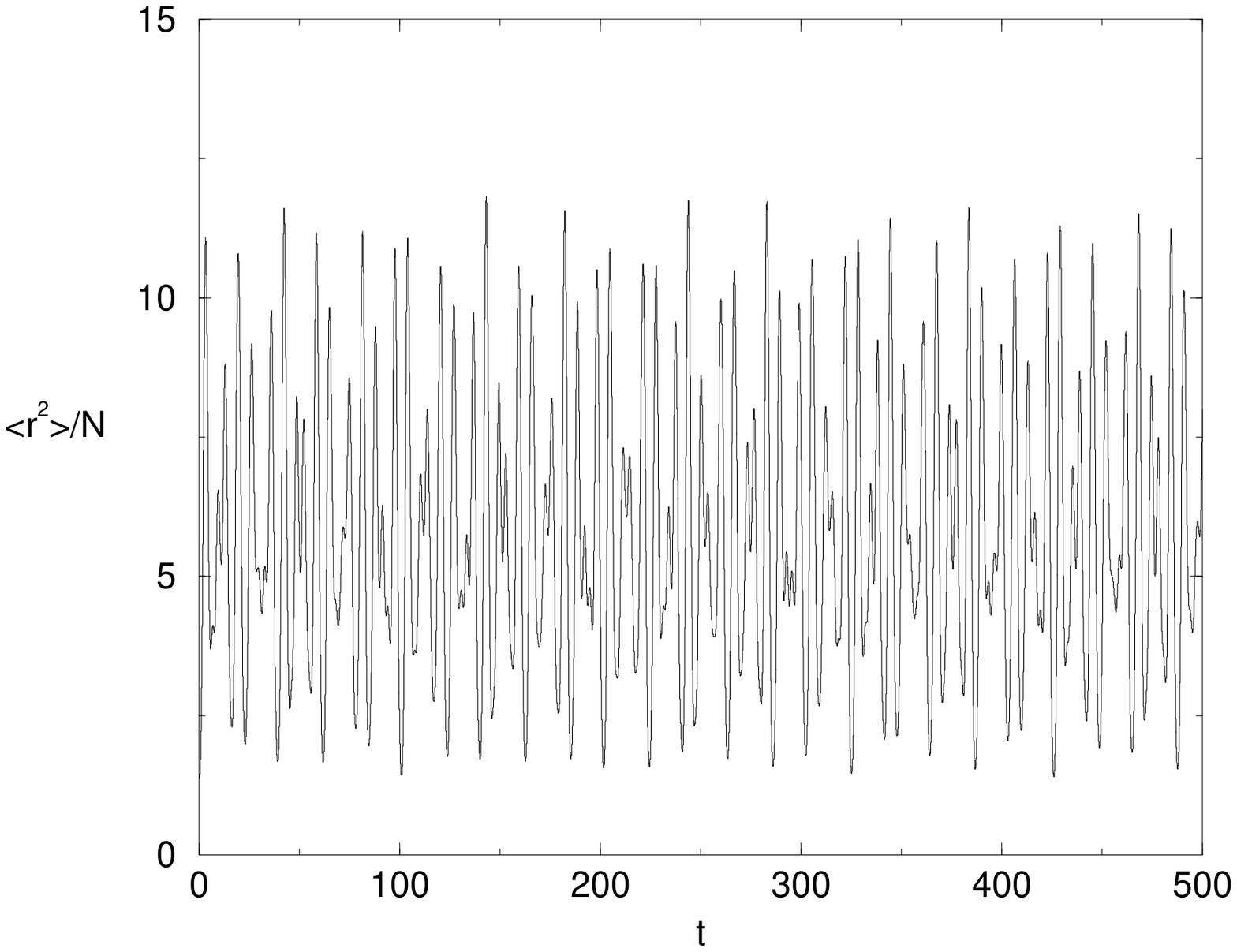,width=3.35in,height=1.5in}}
             {\epsfig{figure=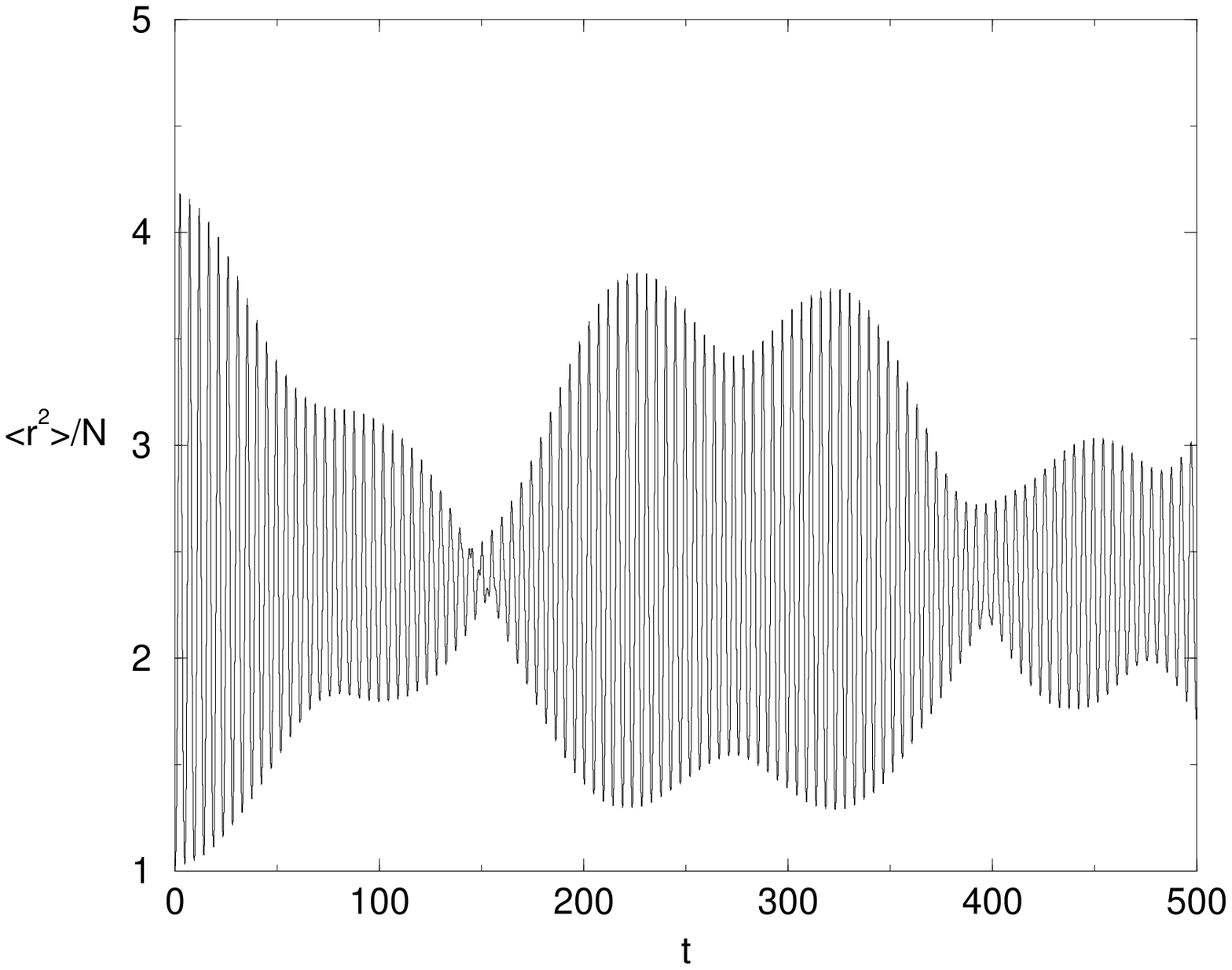,width=3.35in,height=1.5in}}
             \caption{Very long-time behavior of the exact results for
             $\langle r^2 \rangle / N$ for $0 \le t \le 500$. The top
             figure is for $N=1$ and the bottom figure for $N=10$.}
             \label{fig:verylongtime}
\end{figure}

We discussed previously how to choose initial conditions so as to
reach the large $N$ limit in a controlled manner. Starting with these
initial conditions, we show in Fig.~\ref{fig:exact} results from
numerical solution of the Schr\"odinger equation for the time
evolution of $\langle r^2 \rangle / N$. In the strict large $N$ limit
one expects pure harmonic oscillations about the minimum of the
infinite-$N$ effective potential. For any finite value of $N$,
however, at sufficiently large times the pure harmonic motion is
overcome by nonlinear effects, and interesting behavior is found, as
shown in Fig.~\ref{fig:verylongtime}, indicating the presence of a
very non-Gaussian wave function. The ability to capture this long time
behavior is an important test of $1/N$ methods. 

\begin{figure}
   \centering {\epsfig{figure=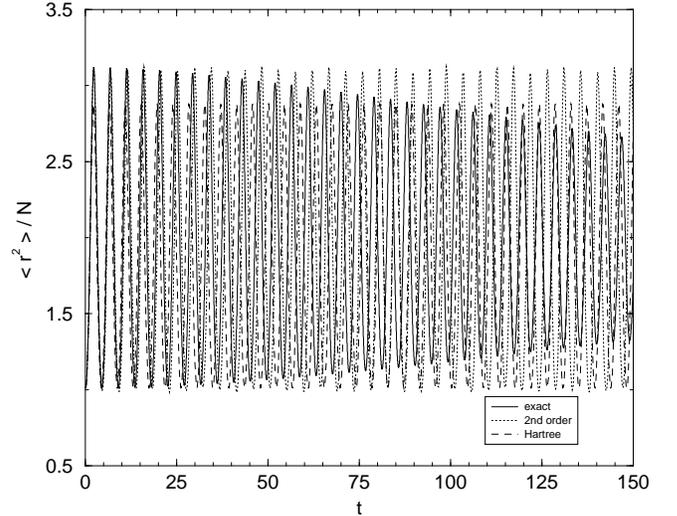,width=3.35in}}
   \caption{$\langle r^2 \rangle / N$ for $N=21$, for the exact,
   next-to-leading order large $N$, and Hartree
   approximations for late times.}  \label{fig:raw_late}
\end{figure}

In order to test whether Hartree or the next-to-leading order
approximation incorporate nonlinearities correctly so as to capture
the late time modulation behavior, we ran a comparison against the
numerical results for $N=21$, the results being displayed in
Fig. (\ref{fig:raw_late}). It is clear that both approximations do not
give satisfactory results. This provides additional motivation for the
development of alternative $1/N$ expansions which would incorporate
selective resummations in order to reduce the coefficient of the error
term at late times.

\subsection{Shifted Gaussian Initial Conditions}

We now discuss the time evolution of a quantum state having an initial
wave function given by Eq.~(\ref{eq:psi0sg}).  For this problem,
because of the lack of symmetry, exact solutions were only obtained
for $N \le 2$.  For $N=1$, depending on whether the energy is above or
below the barrier height, one observes either slow tunneling with
rapid oscillations in one well, or slower oscillation in the complete
range.  At these low values of $N$, the large-$N$ expansion breaks
down quickly, as in the quantum roll, but even here at $N=1$, the
$1/N$ corrections improve the short time accuracy of $q(t)$.

\begin{figure}
   \centering
   \epsfig{figure=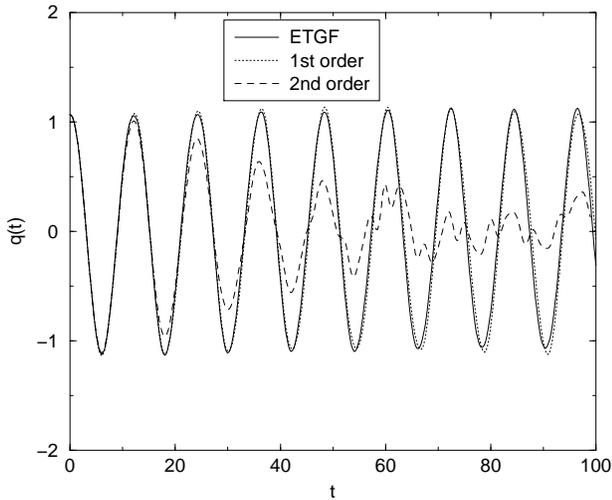,width=3.2in}
   \caption{Plot of $q(t)$ \emph{vs} $t$ for $E > E_b$ with $N=50$.}
\label{fig:dynEgtEb}
\end{figure}

A more relevant comparison is to consider larger values of $N$ at
which the approximations have a better chance of capturing the exact
behavior. In the next figure, we compare the Hartree with the leading
and next-to-leading order large-$N$ approximation for $q(t)$, at
$N=50$. Fig.~\ref{fig:dynEgtEb} displays the results for a run with $E
> E_b$ using the equal time Green's function approximation (see
Sec.~\ref{sec:eqtime}) method for obtaining the Hartree
results. Unlike the situation for the roll initial condition where the
Hartree and next-to-leading order large $N$ results are not
dramatically different, here the qualitative behavior is quite
dissimilar (whereas the Hartree and leading order results are in fact
very close). 

\section{Conclusions}
\label{sec:Cons}

Testing the $1/N$ approximation in quantum mechanics has already
enabled us to arrive at some useful conclusions. In order to interpret
our results, it is important to keep in mind that $1/N$ approximations
are a form of resummed perturbation theory and are therefore only
valid at weak coupling. Thus for couplings of order unity, it is
unrealistic to expect the approximation to give good results for small
values of $N$. Our results have shown that at sufficiently large $N$
the next-to-leading order approximation is a clear improvement over
the leading order approximation, however, at late times this
approximation (as well as Hartree) fails to capture the nonlinear
effects that lead to nontrivial amplitude modulation of the radial
oscillations in the quantum roll problem.

We have noted the presence of a finite-time breakdown in the evolution
given by the next-to-leading order approximation. This result is
related to the fact that the large $N$ expansion for the expectation
values does not necessarily correspond to a positive semi-definite
density matrix when truncated at any finite order in $1/N$.  (At
lowest order, the $1/N$ approximation is equivalent to a Gaussian
variational ansatz for the density matrix and does not have this
problem.)  This last aspect is already clear even in static situations
such as the lack of a real effective potential for all $N$ in the
next-to-leading order approximation. This type of finite-time
breakdown induced by unitarity/positivity violation has also been
noted in simulations of quantum systems where the coupled equal times
Green's function approach was truncated at fourth order \cite{ref:bet}
or where high order cumulant expansion methods were used \cite{shks}.

Two aspects of this breakdown deserve further mention: First, the time
at which breakdown occurs appears to be strongly connected with the
behavior of the effective potential. For values of $N$ not very much
bigger than the critical value $N_c$ (beyond which the effective
potential exits over the entire range of $y$), the breakdown time
increases extremely steeply and may even be pushed to times long
enough to be no longer an obstacle to practical calculations ( this
still needs to be demonstrated). Second, it is important to point out
that avoiding the breakdown via a partial resummation does not
automatically guarantee better late time accuracy (or convergence)
since such a scheme is also only next-to-leading order
accurate. However, it will help in the sense that one may carry out
simulations at smaller values of $N$, thus making it easier to compare
against the late-time numerical solutions of the corresponding
Schr\"odinger equation.

One possible way of correcting the problem of a manifestly positive
operator such as $\langle r^2 \rangle$ becoming negative is to solve
for the full Green's function ${\cal G}_{ij}(t,t')$:
\begin{eqnarray}
   &&
   {\cal G}_{ij}(t,t')
   \ = \
   G_{ij}(t,t') \label{eq:calGfull}  \\
   &&
   - \, \sum_{k,l}
      \int_{\C} {\rm d}t_1 \, \int_{\C} {\rm d}t_2 \,
      G_{ik}(t,t_1) \, \Sigma_{kl}(t_1,t_2) \, {\cal G}_{lj}(t_2,t')
      \>,
   \nonumber 
\end{eqnarray}
rather than the next-to-leading order one as in (\ref{eq:Gfull}).
This equation is the exact equation one obtains by varying the
effective action and it contains terms of all orders in $1/N$ (thus,
strictly speaking, one is no longer truncating at some fixed order).

However, just making this correction does not increase the time period
during which the approximation is accurate.  In order to extend the
accuracy of the $1/N$ approximation to late times, it appears
necessary to use a more robust approximation based on the
Schwinger-Dyson equations.  Several approximations of this sort are
possible, which may both cure the positivity problem as well as lead
to accurate results at late times.  These will be discussed separately
\cite{ref:MDCwork}.

\section{Acknowledgements}

The authors acknowledge helpful conversations with Luis Bettencourt,
Yuval Kluger, and Emil Mottola. S.H. acknowledges stimulating
discussions with Larry Yaffe. The work of B.M. and J.F.D. at UNH is
supported in part by the U.S. Department of Energy under grant
DE-FG02-88ER40410.  B.M. and J.F.D thank the Theory Group (T-8) at
LANL and the Institute for Nuclear Theory at the University of
Washington for hospitality during the course of this work.  F.C. would
like to thank the Physics Department at Yale University for their
hospitality where some of this research was carried out.

\end{document}